\documentclass[letterpaper,10 pt,conference,dvipsnames]{ieeeconf}
\IEEEoverridecommandlockouts

\let\labelindent\relax

\usepackage[absolute,overlay]{textpos}
\usepackage{times}
\usepackage{enumitem}

\usepackage[usenames,dvipsnames]{xcolor}

\usepackage{amsmath,amssymb,mathtools,nccmath, stmaryrd}

\usepackage{amsthm}
\usepackage{bm}
\usepackage{bbold,dsfont,bbm}

\usepackage{graphicx}
\usepackage{tikz,tikz-cd,pgfplots}
\pgfplotsset{compat=1.17}

\usepackage[hidelinks]{hyperref}
\usepackage[capitalise,noabbrev]{cleveref}
\usepackage{cite}

\usepackage{multirow}
\usepackage{booktabs}
\usepackage[acronym]{glossaries}

\usepackage{comment}

\usepackage{units}
\usepackage[per-mode=symbol]{siunitx}[=v2]
\DeclareSIUnit{\eur}{\euro}
\DeclareSIUnit{\usd}{USD}
\DeclareSIUnit{\mph}{mph}
\DeclareSIUnit{\month}{month}
\DeclareSIUnit{\year}{year}
\DeclareSIUnit{\million}{Mil}
\DeclareSIUnit{\mile}{mile}
\DeclareSIUnit{\car}{car}
\DeclareSIUnit{\train}{train}
\DeclareSIUnit{\mmveh}{\text{$\mu$}MV}
\DeclareSIUnit{\nounit}{-}
\usepackage[font=footnotesize]{caption}
\usepackage{subcaption}

\usepackage{array}
\newcommand{\PreserveBackslash}[1]{\let\temp=\\#1\let\\=\temp}
\newcolumntype{C}[1]{>{\PreserveBackslash\centering}p{#1}}
\newcolumntype{R}[1]{>{\PreserveBackslash\raggedleft}p{#1}}
\newcolumntype{L}[1]{>{\PreserveBackslash\raggedright}p{#1}}

\definecolor{lightblue}{rgb}{0.60784,0.76078,0.90196}
\definecolor{darkblue}{rgb}{0.26667,0.44706,0.76863}
\definecolor{lightgreen}{rgb}{0.66275,0.81569,0.55686}
\definecolor{darkgreen}{rgb}{0.43922,0.67843,0.27843}
\definecolor{orange}{rgb}{0.92941,0.49020,0.19216}
\definecolor{yellow}{rgb}{1.00000,0.75294,0.00000}
\definecolor{grey}{rgb}{0.64706,0.64706,0.64706}
\definecolor{purple}{rgb}{0.51373,0.23529,0.04706}
\newacronym{abk:cpo}{CPO}{complete partial order}
\newacronym{abk:cdp}{CDP}{co-design problem}
\newacronym{abk:cdpi}{CDPI}{co-design problem with implementation}
\newacronym{abk:dp}{DP}{design problem}
\newacronym{abk:dpi}{DPI}{design problem with implementation}
\newacronym{abk:mdpi}{MDPI}{monotone design problem with implementation}
\newacronym{abk:dcpo}{DCPO}{directed complete partial order}
\newacronym{abk:mcdp}{MCDP}{Monotone Co-Design Problem}
\newacronym{abk:mcmc}{MCMC}{Markov chain Monte-Carlo}
\newacronym{abk:poset}{poset}{partially ordered set}
\newacronym{abk:uav}{UAV}{unmanned aerial vehicle}
\newacronym{abk:vi}{VI}{variational inference}

\newcommand{\impi}{i}

\newcommand{\prov}{\mathsf{prov}}

\newcommand{\req}{\mathsf{reqs}}

\newcommand{\F}[1]{\textcolor{dpgreen}{#1}}
\newcommand{\FI}[1]{\F{\textit{#1}}}
\newcommand{\funPosetF}{{\F{F}}}

\definecolor{dpgreen}{rgb}{0.0, 0.5, 0.0}

\definecolor{dpred}{rgb}{0.7, 0.0, 0.0}
\newcommand{\R}[1]{\textcolor{dpred}{#1}}
\newcommand{\RI}[1]{\R{\textit{#1}}}
\newcommand{\resPosetR}{{\R{R}}}

\newcommand{\posReals}{\mathbb{R}_{\geq 0}}
\ifPDFTeX

\else

\fi
\newcommand{\sigAlg}{\sigma}
\newcommand{\sigAlgOf}[1]{\sigAlg(#1)}
\newcommand{\sigAlgProduct}{\otimes}
\newcommand{\defeq}{\mathrel{\raisebox{-0.3ex}{$\overset{\text{\tiny def}}{=}$}}}
\newcommand{\setWithArg}[2]{\{#1 \mid #2\}}
\newcommand{\setWithIndex}[2]{\{#1\}_{#2}}

\newcommand{\USet}{\mathtt{U}}
\newcommand{\USetOf}[1]{\USet(#1)}
\newcommand{\upperClosure}[1]{\uparrow\!#1}
\newcommand{\upperOf}[1]{\upperClosure{\{#1\}}}

\newcommand{\DP}{\mathsf{DP}}
\newcommand{\dpOf}[2]{\DP\{#1, #2\}}
\newcommand{\dprb}{{\mathrm{dp}}}
\newcommand{\dprbOf}[1]{\dprb_#1}
\newcommand{\dprba}{\dprbOf{a}}
\newcommand{\dprbb}{\dprbOf{b}}
\newcommand{\dprbL}{\dprbOf{\mathrm{L}}}
\newcommand{\dprbU}{\dprbOf{\mathrm{U}}}
\newcommand{\dprbtask}{\mathrm{task}}

\newcommand{\setproduct}{\times}
\newcommand{\posetproduct}{\times}

\newcommand{\mapFollowing}{\circ}

\newcommand{\mthen}{\fatsemi}
\newcommand{\stack}{\otimes}
\newcommand{\trace}{\text{Tr}}
\newcommand{\traceOf}[1]{\trace(#1)}

\newcommand{\union}{\vee}
\newcommand{\intersection}{\wedge}

\newcommand{\ParaSty}[1]{\tilde{#1}}
\newcommand{\mthenPara}{\mathbin{\ParaSty{\mthen}}}

\newcommand{\mthenUPara}{\mathbin{\bar{\mthen}}}

\newcommand{\IntervalSty}[1]{{#1}_{\Interval}}
\newcommand{\unionInterval}{\mathbin{\IntervalSty{\union}}}

\newcommand{\DistriSty}[1]{{#1}_{\Distri}}
\newcommand{\unionDistri}{\mathbin{\DistriSty{\union}}}

\newcommand{\markovSty}[1]{\hat{#1}}
\newcommand{\MarkovMeasMapf}{\markovSty{\measMapf}}
\newcommand{\mthenMarkov}{\mathbin{\markovSty{\mthen}}}

\newcommand{\MarkovArr}{\rightharpoonup}
\newcommand{\MarkovArrOf}[1]{\overset{#1}{\MarkovArr}}

\newcommand{\mapArrOf}[1]{\overset{#1}{\to}}

\newcommand{\op}{^{\mathrm{op}}}
\newcommand{\inv}{^{-1}}

\newcommand{\interval}[2]{[#1, #2]}

\newcommand{\measure}[1]{\mathbb{#1}}

\newcommand{\measMapf}{f}

\newcommand{\deltaDistOf}[1]{\delta(#1)}
\newcommand{\measureP}{\measure{P}}
\newcommand{\measureQ}{\measure{Q}}
\newcommand{\subsetXP}{X_\posetP}
\newcommand{\subsetYQ}{Y_\posetQ}

\newcommand{\Distri}{\mathcal{D}}
\newcommand{\Interval}{\mathcal{I}}

\newcommand{\DistriOf}[1]{\Distri(#1)}
\newcommand{\IntervalOf}[1]{\Interval(#1)}

\newcommand{\SetOfMap}[2]{{#2}^{#1}}

\newcommand{\SetA}{A}
\newcommand{\SetB}{B}
\newcommand{\SetC}{C}
\newcommand{\SetD}{D}
\newcommand{\SetI}{I}
\newcommand{\SetTB}{{T_B}}

\newcommand{\seteld}{d}
\newcommand{\setelmA}{m_\SetA}
\newcommand{\setelmB}{m_\SetB}
\newcommand{\seteltB}{t_B}

\newcommand{\mapa}{a}
\newcommand{\mapb}{b}
\newcommand{\mapf}{f}
\newcommand{\mapg}{g}
\newcommand{\maph}{h}
\newcommand{\mapr}{r}

\newcommand{\SigAlg}{\Sigma}
\newcommand{\SigAlgOf}[1]{\SigAlg_{#1}}
\newcommand{\SigAlgA}{\SigAlgOf{\SetA}}
\newcommand{\SigAlgB}{\SigAlgOf{\SetB}}

\newcommand{\measureSpaceA}{\tup{\SetA, \SigAlgA}}
\newcommand{\measureSpaceB}{\tup{\SetB, \SigAlgB}}

\newcommand{\measureProduct}{\times}

\newcommand{\subsetYB}{Y_\SetB}
\newcommand{\subsetZC}{Z_\SetC}

\newcommand{\measureKernela}{a}
\newcommand{\measureKernelb}{b}

\newcommand{\measureKernelf}{f}
\newcommand{\measureKernelg}{g}
\newcommand{\measureKernelr}{r}
\newcommand{\measureKernelApply}[2]{(#1 \mid #2)}
\newcommand{\measureKernelfxY}[2]{\measureKernelf\measureKernelApply{#1}{#2}}

\newcommand{\measureKernelSys}{a_\text{Sys}}
\newcommand{\measureKernelUAV}{a_\text{UAV}}

\newcommand{\kernelFollowing}{\circ}

\newcommand{\posetP}{P}
\newcommand{\posetQ}{Q}
\newcommand{\posetR}{R}
\newcommand{\tup}[1]{\langle #1 \rangle}

\newcommand{\posetleq}{\preceq}
\newcommand{\posetgeq}{\succeq}

\newcommand{\SetU}{U}

\newcommand{\poselxF}{x_\funPosetF}

\newcommand{\poselxR}{x_\resPosetR}

\newcommand{\poselxP}{x_\posetP}
\newcommand{\poselyP}{y_\posetP}
\newcommand{\poselxQ}{x_\posetQ}
\newcommand{\poselyQ}{y_\posetQ}
\newcommand{\poselxRnoRes}{x_\posetR}

\definecolor{baiocchi}{RGB}{193,221,245}

\usetikzlibrary{positioning,intersections, hobby, patterns, calc, decorations.pathmorphing, decorations.markings, shadows,shapes, cd,arrows.meta, fit,quotes}

\tikzset{
   tick/.style={postaction={
      decorate,
      decoration={markings, mark=at position 0.5 with {\draw[-] (0,.4ex) -- (0,-.4ex);}}}
   }
}
\tikzstyle{block} = [draw, rectangle, minimum height=2em, minimum width=3em,font=\bfseries,rounded corners,thick]
\tikzstyle{block} = [draw, rectangle, minimum height=2em, minimum width=3em]
\tikzstyle{block1} = [draw, rectangle, minimum height=1.5em, minimum width=2.5em]
\tikzstyle{blockDyn} = [draw, rectangle, minimum height=2.5em, minimum width=3.5em, align=center, inner sep=10pt, thick, fill=white, copy shadow={draw=black,fill=black,opacity=1,shadow xshift=0.5ex,shadow yshift=-0.5ex}]
\tikzstyle{blockAlg} = [draw, rectangle, minimum height=1.5em, minimum width=2.5em, align=center, inner sep=10pt, thick]
\tikzstyle{sum} = [draw,circle]

\tikzstyle{nodePre} = [circle, draw,inner sep=1pt,node contents={$\preceq$},thick]
\tikzstyle{nodePreEmpty} = [circle, draw,inner sep=1pt,thick]
\tikzstyle{nodePos} = [circle, draw,inner sep=1pt,node contents={$\posceq$},thick]
\tikzstyle{nodeProd} = [rectangle, draw,inner sep=4pt,node contents={$\times$},rounded corners,thick]
\tikzstyle{nodeSum} = [rectangle, draw,inner sep=4pt,node contents={$\mathbf{+}$},rounded corners,thick]

\definecolor{red}{rgb}{0.75, 0.0, 0.0}

\tikzset{fcname/.store in =\fcname, fcname={}}
\tikzset{funame/.store in =\funame, funame={}}
\tikzset{rcname/.store in =\rcname, rcname={}}
\tikzset{runame/.store in =\runame, runame={}}
\tikzset{whereres/.store in =\whereres, whereres=0.5}
\tikzset{wherefun/.store in =\wherefun, wherefun=0.5}
\tikzset{relres/.store in =\relres, relres={above}}
\tikzset{relfun/.store in =\relfun, relfun={above}}
\tikzset{posres/.store in =\posres, posres=1}
\tikzset{posfun/.store in =\posfun, posfun=1}
\tikzset{loos/.store in =\loos, loos=2}
\tikzset{feedback/.store in =\feedback, feedback=0}
\tikzset{
   DP/.style={
      label/.style={
         font=\everymath\expandafter{\the\everymath\scriptstyle},
         inner sep=5pt,
         node distance=2pt and -2pt},
      semithick,
      node distance=1 and 1,
      rconn/.style={color=white,opacity=0.0,postaction={decorate}, shorten <=3.2pt, shorten >= 0.8,
      decoration={markings, 
      mark= at position 0 with {
               \coordinate (a);
      },
      mark=at position .5 with
      {
              \ifthenelse{\equal{\feedback}{1}}{\def\angleOut{90}\def\angleIn{90}}{\def\angleOut{0}\def\angleIn{180}}    
              \coordinate (b);
              \draw[dashed,dpred,opacity=1.0] (a) to[out=\angleOut,in=\angleIn,looseness=\loos] 
              node[pos=\posres,\relres=\whereres mm,dpred,opacity=1,fill=white,inner sep=1pt,outer sep=1pt]{\footnotesize{\rcname}} (b);
      },
      mark= at position 1 with 
      {
             \ifthenelse{\equal{\feedback}{1}}{\def\angleOut{0}\def\angleIn{0}}{\def\angleOut{180}\def\angleIn{0}} 
              \ifthenelse{\equal{\feedback}{1}}{\def\symbol{\succeq}}{\def\symbol{\preceq}} 
              \coordinate (c);
              \draw[dpgreen,opacity=1.0] (c) to[out=\angleOut,in=\angleIn,looseness=\loos]
              node[pos=\posfun,\relfun=\wherefun mm,dpgreen,opacity=1,fill=white,inner sep=1pt,outer sep=1pt]{\footnotesize{\fcname}} (b){}; 
              \node[draw,circle,inner sep=0.5pt,color=black,fill=white,opacity=1.0] at (b) (nodepreceq) {$\symbol$}; 
      }
      }},
      runconn/.style={color=dpred,dashed,postaction={decorate},
      decoration={markings,
      mark= at position 1 with {
              \coordinate (a);
              \draw[dpred,opacity=1.0,dashed] ($(a)+(0.05,0)$) --++ (0.5,0) node[\relres,pos=\posres]{\footnotesize{\runame}};}
      }
      },
      funconn/.style={color=white,postaction={decorate},
      decoration={markings,
      mark= at position 0 with {
      \coordinate (a);
      \draw[dpgreen] ($(a)+(-0.05,0)$) -- ($(a)+(-0.5,0)$) node[\relfun, pos=\posfun]{\footnotesize{\funame}};}
      }
      },
      execute at begin picture={\tikzset{
         x=\dpx, y=\dpy,
         every fit/.style={inner xsep=\dpx, inner ysep=\dpy}}}
      },
   dpx/.store in=\dpx,
   dpx = 1.5cm,
   dpy/.store in=\dpy,
   dpy = 1.5ex,
   dp port sep/.store in=\dpportsep,
   dp port sep=2,
   dp port length/.store in=\dpportlen,
   dp port length=4pt,
   dp min width/.store in=\dpminwidth,
   dp min width=0.5cm,
   dp rounded corners/.store in=\dpcorners,
   dp rounded corners=2pt,
   dp small/.style={dp port sep=1, dp port length=2.5pt, dpx=.4cm, dp min width=.4cm, dpy=.7ex},
   dp/.code 2 args={
      \pgfmathsetlengthmacro{\dpheight}{\dpportsep * (max(#1,#2)) * \dpy}
      \pgfkeysalso{draw,%
        minimum width=\dpminwidth,%
        minimum height=\dpheight,%
        font=\bfseries,
        outer sep=0pt,%
        inner sep=5pt,%
        rounded corners=\dpcorners,
        thick,
        prefix after command={\pgfextra{\let\fixname\tikzlastnode}},
        append after command={\pgfextra{\draw
            \ifnum #1=0{} \else foreach \i in {1,...,#1} { 
            ($(\fixname.north west)!{\i/(#1+1)}!(\fixname.south west)$) +(0,0) node[solid,left,circle,color=dpgreen,draw,fill=dpgreen,scale=0.3] {} coordinate (\fixname_fun\i) -- +(0,0) coordinate (\fixname_fun\i')}\fi 
            \ifnum #2=0{} \else foreach \i in {1,...,#2} {
            ($(\fixname.north east)!{\i/(#2+1)}!(\fixname.south east)$) +(0,0) coordinate (\fixname_res\i') -- +(0,0) node[solid,right,circle,color=dpred,draw,fill=dpred,scale=0.3] {} coordinate (\fixname_res\i)}\fi;
         }}}
         },
      dp name/.style={append after command={\pgfextra{\node[label=center,inner sep=2pt,fill=white] at (\fixname) {\textbf{#1}};}}}
   }
\renewcommand{\baselinestretch}{0.985}
\theoremstyle{definition}

\newtheorem*{assumption*}{Assumption}
\newtheorem{theorem}{Theorem}

\newtheorem{lemma}[theorem]{Lemma}

\newtheorem{proposition}[theorem]{Proposition}
\newtheorem{definition}{Definition}
\theoremstyle{remark}

\Crefname{figure}{Fig.}{Figures}
\crefname{equation}{Eq.}{Eqs.}
\makeatletter
    \if@cref@capitalise
        \crefname{subsection}{Section}{Sections}
        \crefname{subsubsection}{Section}{Sections}
        \crefname{assumption}{Assumption}{Assumptions}
        \crefname{problem}{Problem}{Problems}
    \else
        \crefname{subsection}{section}{sections}
        \crefname{subsubsection}{section}{sections}
        \crefname{assumption}{assumption}{assumptions}
        \crefname{problem}{problem}{problems}
    \fi
\makeatother

\usetikzlibrary{positioning,intersections, hobby, patterns, calc, decorations.pathmorphing, decorations.markings, shadows,shapes, cd,arrows.meta, fit,quotes}
\tikzset{
   tick/.style={postaction={
      decorate,
      decoration={markings, mark=at position 0.5 with {\draw[-] (0,.4ex) -- (0,-.4ex);}}}
   }
}
\tikzstyle{block} = [draw, rectangle, minimum height=2em, minimum width=3em,font=\bfseries,rounded corners,thick]
\tikzstyle{block1} = [draw, rectangle, minimum height=1.5em, minimum width=2.5em]
\tikzstyle{blockDyn} = [draw, rectangle, minimum height=2.5em, minimum width=3.5em, align=center, inner sep=10pt, thick, fill=white, copy shadow={draw=black,fill=black,opacity=1,shadow xshift=0.5ex,shadow yshift=-0.5ex}]
\tikzstyle{blockAlg} = [draw, rectangle, minimum height=1.5em, minimum width=2.5em, align=center, inner sep=10pt, thick]
\tikzstyle{sum} = [draw,circle]

\tikzstyle{blockfill} = [block,rounded corners=4,fill=white]

\tikzstyle{nodePre} = [circle, draw,inner sep=1pt,node contents={$\preceq$},thick]
\tikzstyle{nodePreEmpty} = [circle, draw,inner sep=1pt,thick]
\tikzstyle{nodePos} = [circle, draw,inner sep=1pt,node contents={$\posceq$},thick]
\tikzstyle{nodeProd} = [rectangle, draw,inner sep=4pt,node contents={$\times$},rounded corners,thick]
\tikzstyle{nodeSum} = [rectangle, draw,inner sep=4pt,node contents={$\mathbf{+}$},rounded corners,thick]

\definecolor{DPgreen}{rgb}{0.0, 0.5, 0.0}
\definecolor{red}{rgb}{0.75, 0.0, 0.0}

\newif\ifmargincomments 
\margincommentstrue

\newif\ifextendedversion 

\ifmargincomments

\else

\fi

\title{
\textbf{On Composable and Parametric Uncertainty in Systems Co-Design}
}
\author{Yujun Huang$^1$, Marius Furter$^2$, Gioele Zardini$^1$
\thanks{
$^1$Laboratory for Information and Decision Systems, Massachusetts Institute of Technology, Cambridge (MA), USA, {\tt \{yujun233,gzardini\}@mit.edu}}
\thanks{
$^2$Department of Mathematics, University of Zurich, Switzerland, \tt{marius.furter@math.uzh.ch}
}
}

\begin{document}


\maketitle

\begin{abstract}
Optimizing the design of complex systems requires navigating interdependent decisions, heterogeneous components, and multiple objectives.
Our monotone theory of co-design offers a compositional framework for addressing this challenge, modeling systems as \glspl{abk:dp}, representing trade-offs between functionalities and resources within partially ordered sets.
While current approaches model uncertainty using intervals, capturing worst- and best-case bounds, they fail to express probabilistic notions such as risk and confidence.
These limitations hinder the applicability of co-design in domains where uncertainty plays a critical role.
In this paper, we introduce a unified framework for \emph{composable uncertainty} in co-design, capturing intervals, distributions, and parametrized models.
This extension enables reasoning about risk-performance trade-offs and supports advanced queries such as experiment design, learning, and multi-stage decision making.
We demonstrate the expressiveness and utility of the framework via a numerical case study on the uncertainty-aware co-design of task-driven \glspl{abk:uav}.
\end{abstract}

\glsresetall


\section{Introduction}
Designing embodied systems involves complex trade-offs between hardware components, such as sensors, actuators, and processors, and software modules for perception, planning, and control~\cite{merlet2005optimal,zhu2018codesign,saoud2021compositional,shell2021design,zardiniCoDesignComplexSystems2023}.
Traditional methods often optimize subsystems in isolation, limiting both modularity and interdisciplinary collaboration~\cite{seshia2016design,zardiniCoDesignComplexSystems2023}.
Over the past few years, we have introduced a monotone framework for co-design, which allows one to formulate and solve complex, compositional design optimization problems leveraging domain theory and category theory~\cite{zardiniCoDesignComplexSystems2023}.
The existing toolbox has been successfully applied to solve problems in robotics and controls~\cite{zardiniecc21,zardiniTaskdrivenModularCodesign2022,milojevic2025codei}, transportation~\cite{zardini2022co}, and automotive~\cite{neumann2024co}.
However, existing approaches model uncertainty only via interval bounds, which guarantee robustness but lack the expressiveness needed to capture risk, probability of success, or adaptive decision-making under uncertainty.
In this work, we extend the co-design framework to handle richer forms of uncertainty, including distributions and parametrized models, enabling queries over probabilistic trade-offs and paving the way for learning, estimation, and adaptive optimization in the design of complex systems.
Specifically, we formalize uncertainty as composable structures over \glspl{abk:dp}, preserving the compositionality of co-design operations.
We illustrate our approach through the co-design of \glspl{abk:uav}, showing how uncertainty-aware design unlocks new capabilities for robust and efficient compositional decision-making.

\paragraph*{Organization of the paper}
The remainder of this paper is organized as follows.
\cref{sec:co-design} reviews the foundations of our monotone co-design theory, including the current approach to uncertainty quantification via intervals.
\cref{sec:uncertain-param-dp} extends the existing theory to model distributional and parameteric uncertainty in \glspl{abk:dp}.
We use the design of payload-carrying \glspl{abk:uav} as a motivating example across this paper, and present a numerical study in \cref{sec:uav-example}.
\Cref{sec:conclusions} concludes the work, with proof sketches in \cref{sec:proofs}.

\section{Monotone Co-Design Theory}
\label{sec:co-design}
After introducing the required preliminaries, we summarize the main concepts of monotone co-design~\cite{Censi2015, censi2022, zardiniCoDesignComplexSystems2023}. 

\subsection{Mathematical preliminaries}\label{subsec:math-preliminary}

\subsubsection{Sets and functions}\label{subsubsec:sets-functions}
We write~$\mapf \colon \SetA  \to \SetB$ for functions between sets $\SetA$ and $\SetB$ and indicate the action of $\mapf$ on elements by $\setelmA \mapsto \mapf(\setelmA)$.
We call $\SetA$ the \emph{domain} of $\mapf$, and $\SetB$ its \emph{co-domain}.
We will often use the broad term \emph{map} to refer to functions.
For a map~$\mapf \colon \SetA \to \SetB$, we denote the pre-image of $\subsetYB \subseteq \SetB$ as the set of elements in $\SetA$ whose image lies in $\subsetYB$, $\mapf\inv\left(\subsetYB\right) \defeq \setWithArg{\setelmA \in \SetA}{\mapf(\setelmA) \in \subsetYB}$.
Given maps $\mapf \colon \SetA \to \SetB$ and $\mapg \colon \SetB \to \SetC$, their \emph{composite}
is the map $\mapg \mapFollowing \mapf \colon \SetA \to \SetC$ that sends $ \setelmA \mapsto \mapg(\mapf(\setelmA))$. We will often express composition diagrammatically:~$
\SetA \mapArrOf{\mapf} \SetB \mapArrOf{\mapg} \SetC$.
We write $\SetA \setproduct \SetB$ for the \emph{Cartesian product} of sets. Its elements are tuples $\tup{\setelmA, \setelmB}$, where $\setelmA \in \SetA$ and $\setelmB \in \SetB$.
Given maps $\mapf \colon \SetA \to \SetB$ and $\mapg \colon \SetA' \to \SetB'$, their \emph{product} is
\begin{equation*}
    \begin{aligned}
        \mapf \setproduct \mapg \colon \SetA \setproduct \SetA'  & \to \SetB \setproduct \SetB', \\
        \tup{\setelmA, \setelmA'} &\mapsto \tup{\mapf(\setelmA), \mapg(\setelmA')}.
    \end{aligned}
\end{equation*}
Given $\maph \colon \SetA \setproduct \SetB \to \SetC$, we denote its \emph{partial evaluation} by
\begin{equation*}
    \begin{aligned}
        \maph(-,\setelmB) \colon \SetA  & \to \SetC, \\
        \setelmA &\mapsto \maph(\setelmA, \setelmB).
    \end{aligned}
\end{equation*}

\subsubsection{Background on orders}\label{sec:app_order}

\begin{definition}[Poset]
A \emph{\gls{abk:poset}} is a tuple $\mathcal{P} =\tup{ P,\preceq_\mathcal{P}}$, where $P$ is a set and~$\preceq_\mathcal{P}$ is a partial order (a reflexive, transitive, and antisymmetric relation). 
If clear from context, we use~$\posetP$ for a \gls{abk:poset}, and~$\posetleq$ for its order.
\end{definition}

\begin{definition}[Opposite poset]
The \emph{opposite} of a \gls{abk:poset}~$\mathcal{P} = \tup{ P,\preceq_\mathcal{P}}$ is the poset $\mathcal{P}\op \defeq\tup{ P,\posetleq_{\mathcal{P}}\op }$ with the same elements and reversed ordering:
$
\poselxP \posetleq_{\mathcal{P}}\op \poselyP \Leftrightarrow 
\poselyP \posetleq_{\mathcal{P}}\poselxP
$.
\end{definition}

\begin{definition}[Product poset]
Given \glspl{abk:poset} $\tup{P,\preceq_{\mathcal{P}}}$ and $\tup{Q,\preceq_{\mathcal{Q}}}$, their \emph{product} $\tup{P\times Q,\preceq_{\mathcal{P}\times \mathcal{Q}}}$ is the poset with
\begin{equation*}
    \tup{\poselxP,\poselxQ}\preceq_{\mathcal{P}\times \mathcal{Q}}\tup{\poselyP,\poselyQ} \Leftrightarrow (\poselxP \preceq_{\mathcal{P}} \poselyP) \wedge (\poselxQ \preceq_\mathcal{Q} \poselyQ).
\end{equation*}
\end{definition}

\begin{definition}[Upper closure]\label{def:upper-losure}
    Let $\posetP$ be a \gls{abk:poset}. The \emph{upper closure} of a subset $\subsetXP \subseteq \posetP$ contains all elements of $\posetP$ that are greater or equal to some $\poselyP \in \subsetXP$:
    \begin{equation*}
        \upperClosure{\subsetXP} \defeq \setWithArg{\poselxP \in \posetP}{\exists \poselyP \in \subsetXP : \poselyP \posetleq_\posetP \poselxP}.
    \end{equation*}
\end{definition}

\begin{definition}[Upper set]\label{def:uppersets-of-posets}
    A subset $\subsetXP \subseteq \posetP$ of a \gls{abk:poset} is called an \emph{upper set} if it is upwards closed: $\upperClosure{\subsetXP} = \subsetXP$.
    We write $\USetOf{\posetP}$ for the set of upper sets of $\posetP$.
    We regard $\USetOf{\posetP}$ as partially ordered under $\SetU \posetleq \SetU' \Leftrightarrow \SetU \supseteq \SetU'$.
\end{definition}


\begin{definition}[Monotone map]
A map $f\colon \posetP \to \posetQ$ between \glspl{abk:poset} $\langle P, \preceq_\mathcal{P} \rangle$ and $\langle Q, \preceq_\mathcal{Q} \rangle$ is  \emph{monotone} if $x\preceq_\mathcal{P} y$ implies $f(x) \preceq_\mathcal{Q} f(y)$. Monotonicity is preserved by composition and products.
\end{definition}


\subsubsection{Background on probability}\label{subsubsec:sigma-alg-intro}

\begin{definition}[Measurable space]
    A \emph{sigma algebra} $\SigAlgA$ on set $\SetA$ is a non-empty collection of subsets of $\SetA$ that is closed under complements, countable unions, and countable intersections. The tuple $\measureSpaceA$ is called a \emph{measurable space}. For an arbitrary collection of subsets $\mathcal{G}$, the sigma algebra $\sigAlgOf{\mathcal{G}}$ \emph{generated} by $\mathcal{G}$ is the smallest sigma algebra containing $\mathcal{G}$.
    Given another measurable space $\measureSpaceB$, a map $\mapf \colon \SetA \to \SetB$ is called \emph{measurable} if pre-images of measurable sets are measurable: $\mapf^{-1}(Y) \in \SigAlgA , \ \forall Y \in \SigAlgB$. 
\end{definition}

\begin{definition}[Probability distribution]
    A \emph{probability distribution} on a measurable space $\measureSpaceA$ is a map $\measureP \colon \SigAlgA \to \interval{0}{1}$ satisfying $\measureP(\SetA) = 1$ and
    $\measureP(\bigcup_{i=1}^\infty E_i) = \sum_{i=1}^\infty \measureP(E_i)$ for any collection $\{E_i\}_{i=1}^\infty$ of disjoint sets.
    Elements in $\SetA$ are called \emph{outcomes} or \emph{results} and sets in the sigma algebra are called \emph{events}.
    $\measureP(E)$ denotes the \emph{probability} of event $E$.
\end{definition}

\subsection{Monotone co-design theory}\label{subsec:co-design-intro}
Co-design provides a compositional framework for the analysis of complex systems.
Consider the design of \glspl{abk:uav}, where perception is an important sub-system.
With other conditions fixed, the perception sub-system can be viewed as providing a certain level of \FI{detection accuracy} at the cost of \RI{computation power}, under given \RI{weather conditions}.
While detection accuracy and computation power can be naturally modeled as positive real numbers, weather conditions are complex and involve non-comparable elements.
For instance, a clear night and a foggy day pose qualitatively different challenges to the perception system. 
Hence, designs optimized for one case will not necessarily perform well under the other.
Monotone co-design formalizes such situations as \glspl{abk:dp} which provide \FI{functionalities} at the cost of \RI{resources} (colored by their respective roles), both assumed to be partially ordered.
This enables expressing trade-offs between incomparable optimal designs.

\begin{definition}[\Gls{abk:dp}]
    Given \glspl{abk:poset}~$\funPosetF$ and $\resPosetR$ of \FI{functionalities} and \RI{resources}, a \emph{\gls{abk:dp}} is an upper set of $\funPosetF\op \posetproduct \resPosetR$.
    We denote the set of such \glspl{abk:dp} by $\dpOf{\funPosetF}{\resPosetR}$.
    Given a \gls{abk:dp} $\dprb$, a pair $\tup{\poselxF, \poselxR}$ of functionality $\poselxF$ and resource $\poselxR$ is \emph{feasible} if $\tup{\poselxF, \poselxR} \in \dprb$.
    We order $\dpOf{\funPosetF}{\resPosetR}$ by inclusion: $\dprba \posetleq \dprbb \Leftrightarrow \dprba \subseteq \dprbb$.
    Note that this is the opposite of the ordering used for upper sets.
\end{definition}

The upper set condition captures the following intuition: If resource $\poselxR$ suffices to provide functionality $\poselxF$, then it also suffices for any worse functionality $\poselxF' \posetleq \poselxF$.
Moreover, any better resource $\poselxR' \posetgeq \poselxR$ should also suffice to provide $\poselxF$.

A core tenet of co-design is to compose systems out of simpler sub-systems. Such composites are formalized as multi-graphs of \glspl{abk:dp} called \emph{co-design problems}.
We summarize the main composition operations in \cref{def:dp-interconnections} with some shown diagrammatically in \cref{fig:dp_def}.
An example of co-design diagram for \glspl{abk:uav} can be seen in \cref{fig:uav-dp-composition}, introduced in \cref{sec:uav-example}.

\begin{figure}[tb]
\begin{subfigure}{\columnwidth}
    \begin{center}
    \begin{tikzpicture}[DP]
            \node[dp={2}{2}] (cnt) {DP};
            \draw[runconn, runame={resources}, relres=above,posres=0.9] (cnt_res1){};
            \draw[runconn, runame={}, relres=above,posres=0.9] (cnt_res2){};
            \draw[funconn, funame={functionalities},relfun=above,posfun=1.15] (cnt_fun1){};
            \draw[funconn, funame={},relfun=above,posfun=1.15] (cnt_fun2){};
\end{tikzpicture}
    \subcaption{A \gls{abk:dp} is a monotone relation between posets of \F{functionalities} and \R{resources}. \label{fig:mathcodesign}}
    \end{center}
    \end{subfigure}
\begin{center}
\begin{subfigure}[b]{0.3\columnwidth}
\centering
\scalebox{0.8}{\begin{tikzpicture}[DP]
    \node[dp={1}{1}] (f) {$d$};
    \node[dp={1}{1}, right=1cm of f] (g) {$e$};
    \draw[rconn, rcname={}, fcname={}] (f_res1)  to (g_fun1);
    \draw[runconn, runame={}] (g_res1);
    \draw[funconn, funame={}] (f_fun1);
\end{tikzpicture}}
\subcaption{Series.}
\end{subfigure}
\begin{subfigure}[b]{0.3\columnwidth}
\centering
\scalebox{0.8}{\begin{tikzpicture}[DP]
    \node[dp={1}{1}] (f) {$d$};
    \node[dp={1}{1}, below=0.3cm of f] (g) {$e$};
    \draw[runconn, runame={}] (f_res1){};
    \draw[runconn, runame={}] (g_res1){};
    \draw[funconn, funame={}] (f_fun1){};
    \draw[funconn, funame={}] (g_fun1){};
\end{tikzpicture}}
\subcaption{Parallel.}
\end{subfigure}
\begin{subfigure}[b]{0.3\columnwidth}
\centering
\scalebox{0.8}{\begin{tikzpicture}[DP]
    \node[dp={2}{2}] (f) {$d$};
    \draw[runconn, runame={}] (f_res2){};
    \draw[funconn, funame={}] (f_fun2){};
    \draw[rconn,rcname={},fcname={},feedback=1,loos=3] (f_res1) -- ($(f)+(0,4)$) |- (f_fun1);
\end{tikzpicture}}
\subcaption{Loop.}
\end{subfigure}
\label{fig:diagrams}
\caption{\Glspl{abk:dp} can be composed in different ways.}
\label{fig:dp_def}
\end{center}
\vspace{-5mm}
\end{figure}

\begin{definition}[Composition operations for \glspl{abk:dp}]\label{def:dp-interconnections}
    The following operations construct new \glspl{abk:dp} from old:\\
    \emph{Series}: Given \glspl{abk:dp} $\dprba \in \dpOf{\posetP}{\posetQ}$ and $\dprbb \in \dpOf{\posetQ}{\posetR}$, their series connection $\dprba \mthen \dprbb \in \dpOf{\posetP}{\posetR}$ is defined as
        \[            
        \{\tup{\poselxP, \poselxRnoRes} \mid
            \exists \poselxQ : \tup{\poselxP, \poselxQ} \in \dprba
            \text{ and }\tup{\poselxQ, \poselxRnoRes} \in \dprbb\}.
        \]
    This models situations where $\dprba$ uses the functionalities provided by $\dprbb$ as its resources. \\
    \emph{Parallel}: For $\dprba \in \dpOf{\posetP}{\posetQ}$ and $\dprba' \in \dpOf{\posetP'}{\posetQ'}$, their parallel connection $\dprba \stack \dprba' \in \dpOf{\posetP \posetproduct \posetP'}{\posetQ \posetproduct \posetQ'}$ is
        \[
            \{\tup{\tup{\poselxP,\poselxP'}, \tup{\poselxQ,\poselxQ'}} \mid 
            \tup{\poselxP, \poselxQ} \in \dprba
            , \tup{\poselxP', \poselxQ'} \in \dprba' \}.
        \]
    It represents two non-interacting systems. \\
    \emph{Feedback/Trace}: For $\dprb \in \dpOf{\posetP \posetproduct \posetR}{\posetQ \posetproduct \posetR}$, its trace $\traceOf{\dprb} \in \dpOf{\posetP}{\posetQ}$ is defined as
        \[
            \{ \tup{\poselxP, \poselxQ} \mid
            \exists \poselxRnoRes : \tup{\tup{\poselxP,\poselxRnoRes}, \tup{\poselxQ,\poselxRnoRes}} \in \dprb \}
        \]
    This models the case where functionalities provided by $\dprb$ are used as its own resources. \\
    \emph{Union} and \emph{intersection}: Given $\dprba,\dprbb \in \dpOf{\posetP}{\posetQ}$, their union $\dprba \union \dprbb \in \dpOf{\posetP}{\posetQ}$ is defined by
        \[
            \{\tup{\poselxP,\poselxQ} \mid \\
            \tup{\poselxP, \poselxQ} \in \dprba
            \text{ or }\tup{\poselxP, \poselxQ} \in \dprbb \}.
        \]
    Designing for the union expresses a free choice between satisfying $\dprba$ or $\dprbb$.
    Similarly, the intersection $\dprba \intersection \dprbb \in \dpOf{\posetP}{\posetQ}$ is defined as
        \[
            \{\tup{\poselxP,\poselxQ} \mid
            \tup{\poselxP, \poselxQ} \in \dprba
            \text{ and }\tup{\poselxP, \poselxQ} \in \dprbb \}.
        \]
    Designing for the intersection requires satisfying both $\dprba$ and $\dprbb$. Note that union and intersection can be applied to a set of \glspl{abk:dp}, for instance $\union\setWithIndex{\dprb_i}{i \in I}$.
\end{definition}
%

Designers care not only about the best performance achievable by a system, but also about what design choices realize this optimal value.
To answer such questions, co-design models design choices using \emph{implementations}.

\begin{definition}[\Gls{abk:mdpi}]\label{def:mdpi}
    Given \glspl{abk:poset}~$\funPosetF$ and $\resPosetR$, an \gls{abk:mdpi} consists of a set of \emph{implementations} $\SetI$, along with maps $\prov \colon \SetI \to \funPosetF$ and $\req \colon \SetI \to \resPosetR$.
    For each design choice $\impi \in \SetI$, $\prov(\impi)$ represents the \FI{functionality} provided by $\impi$, while $\req(\impi)$ represents the \RI{resource} it requires.
    For each \gls{abk:mdpi}, there is a corresponding \gls{abk:dp} given by the free choice among all implementations: 
    $    \union\setWithIndex{\upperOf{\tup{\prov(\impi), \req(\impi)}}}{\impi \in \SetI}.
    $
    If a pair $\tup{\poselxF, \poselxR} \in \dprb$ is feasible with respect to this \gls{abk:dp}, then there exists an implementation in $\SetI$ that provides a functionality $\poselxF' \posetgeq \poselxF$ for a resource $\poselxR' \posetleq \poselxR$.
\end{definition}


With these modeling techniques in hand, we can ask for optimal solutions to \glspl{abk:dp}. These are formalized as \emph{queries}.

\begin{definition}[Querying \glspl{abk:dp}]\label{def:queries-dp}
    Given a \gls{abk:dp}~$\dprb \in \dpOf{\funPosetF}{\resPosetR}$, we define two types of queries:
    \begin{enumerate}
        \item \emph{Fix \FI{functionalities} minimize \RI{resources}}: For a fixed $\poselxF \in \funPosetF$, return the set of resources $\poselxR \in \resPosetR$ that make $\tup{\poselxF, \poselxR}$ feasible with respect to $\dprb$. We can view this query as a monotone map $q \colon \funPosetF \to \USetOf{\resPosetR}$.
        \item \emph{Fix \RI{resources} maximize \FI{functionalities}}: For a fixed $\poselxR \in \resPosetR$, return the set of functionalities $\poselxF \in \funPosetF$ that make $\tup{\poselxF, \poselxR}$ feasible with respect to $\dprb$. We can view this query as a monotone map $q' \colon \resPosetR \to \USetOf{\funPosetF\op}$.
    \end{enumerate}
\end{definition}

Since $\funPosetF,\resPosetR$ are \glspl{abk:poset}, computing query results and feasible implementations is a \emph{multi-objective optimization problem}.
Using the compositional structure, one can derive efficient algorithms to calculate the query results for complex \glspl{abk:dp} defined by a multi-graph of sub-systems~\cite{zardiniCoDesignComplexSystems2023}.

\subsection{Interval uncertainty in co-design}\label{subsec:interval-uncertain-dp}

Currently, uncertainty in co-design is modeled via \emph{intervals} of \glspl{abk:dp}~\cite{censi2017uncertainty}.
Recall that we order $\dpOf{\funPosetF}{\resPosetR}$ by inclusion, so a ``better'' \gls{abk:dp} has more resource/functionality pairs feasible.
When facing uncertainty in a system, one can bound its performance with \emph{optimistic} (best case) and \emph{pessimistic} (worst case) systems.

\begin{definition}[Interval uncertainty of \glspl{abk:dp}]
    Given \glspl{abk:poset}~$\funPosetF$ and $\resPosetR$, the set of interval \glspl{abk:dp} is defined as
    \begin{multline*}
        \IntervalOf{\dpOf{\funPosetF}{\resPosetR}} \defeq \\
        \setWithArg{\interval{\dprbL}{\dprbU}}{\dprbL, \, \dprbU \in \dpOf{\funPosetF}{\resPosetR}, \, \dprbL \posetleq \dprbU},
    \end{multline*}
    where the lower bound $\dprbL$ represents the pessimistic estimate, and $\dprbU$ represents the optimistic estimate.
\end{definition}


\begin{lemma} \label{lem:interval-lifted-ops}
    All the operations in \cref{def:dp-interconnections} can be lifted to intervals of \glspl{abk:dp}.
    The lifted operations are
    \begin{align*}
        \IntervalSty{\trace}(\interval{\dprbL}{\dprbU}) &\defeq \interval{\traceOf{\dprbL}}{\traceOf{\dprbU}}, \\
        \interval{\dprbL}{\dprbU} \mathbin{\IntervalSty{\diamond}} \interval{\dprbL'}{\dprbU'} &\defeq \interval{\dprbL \diamond \dprbL'}{\dprbU \diamond \dprbU'},
    \end{align*}
    where~$\diamond$ is either $\mthen$, $\stack$, $\union$ or $\intersection$. 
    Moreover, $\dpOf{\funPosetF}{\resPosetR}$ embeds into $\IntervalOf{\dpOf{\funPosetF}{\resPosetR}}$ by $\dprb \mapsto \interval{\dprb}{\dprb}$~\cite{censi2017uncertainty}.
\end{lemma}
Consequently, we can view multi-graphs like the one in \cref{fig:uav-dp-composition} as representing composites of intervals of \glspl{abk:dp}.
Solving queries for interval uncertainty results in separate results for the optimistic and pessimistic cases \cite{censi2017uncertainty}.
\section{Uncertainty and Parameterization}\label{sec:uncertain-param-dp}

Taking a closer look at the perception system from \cref{subsec:co-design-intro}, one realizes that many state-of-the-art algorithms are sampling-based, yielding performance guarantees only in terms of \emph{probability distributions}.
Manufacturing of sensors is also an uncertain process, resulting in \emph{distributions} over parameters and performance.
Therefore, even for fixed \RI{computation power} and \RI{weather condition}, a given implementation may only guarantee a distribution over the provided \FI{detection accuracy}.
This motivates the need to incorporate distributional uncertainty into the co-design process.

In this section, we describe a new formal, unified language for uncertainty in co-design, which incorporates intervals, subsets, and distributions over \glspl{abk:dp}.
In addition, we introduce parameterization for both \glspl{abk:dp} and uncertain \glspl{abk:dp}, which allow us to express dependencies among design choices and other factors.
In all cases, we show how the composition operations lift to the new structures. 
It follows from the general theory established in~\cite{furter2025composableuncertaintysymmetricmonoidal} that the lifted operations still have desirable compositional properties.

\subsection{Distributional framework for uncertainty in co-design}\label{subsec:uncertain-dp}

To define probability distributions on sets of \glspl{abk:dp}, we exploit the fact that they are partially ordered.

\begin{definition}[Probability distributions on \glspl{abk:poset}]\label{def:prob-dist-poset}
    For a \gls{abk:poset} $\posetP$, consider the sigma algebra generated by $\USetOf{\posetP}$ (the upper sets of $\posetP$), denoted as $\sigAlgOf{\posetP}$.
    We define $\DistriOf{\posetP}$ to be the set of probability distributions on $\tup{\posetP,\sigAlgOf{\posetP}}$.
\end{definition}
The following lemma explains the relationship between posets $\posetP$ and the set $\DistriOf{\posetP}$ of distributions on them. It holds for probability distributions on any space.

\begin{lemma}\label{lem:inclusion-to-dist}
    For every poset~$\posetP$, there is an inclusion map $P \hookrightarrow \DistriOf{\posetP}$ that sends an element~$\poselxP$ to the delta distribution~$\deltaDistOf{\poselxP}$ defined by
    \[
        \deltaDistOf{\poselxP}(\subsetXP) \defeq 
        \begin{cases}
            1, & \text{if } \poselxP \in \subsetXP, \\
            0, & \text{otherwise.}
        \end{cases}
    \]
    Moreover, each measurable map~$\measMapf \colon \posetP \to \posetQ$, lifts to a function $\DistriOf{\measMapf}\colon \DistriOf{\posetP}\to \DistriOf{\posetQ}$ that sends a probability distribution~$\measureP \colon \sigAlgOf{\posetP} \to \interval{0}{1}$ to the distribution $\measureQ \colon \sigAlgOf{\posetQ} \to \interval{0}{1}$ defined by $\measureQ(\subsetYQ) = \measureP(\measMapf\inv(\subsetYQ))$ for each $\subsetYQ$ in $\sigAlgOf{\posetQ}$.
\end{lemma}

Recall that the set of \glspl{abk:dp} $\dpOf{\funPosetF}{\resPosetR}$ is partially ordered by inclusion. Hence, we can apply \cref{def:prob-dist-poset} to $\dpOf{\funPosetF}{\resPosetR}$, obtaining $\DistriOf{\dpOf{\funPosetF}{\resPosetR}}$ as distributions of \glspl{abk:dp}.
To interpret co-design problems in this new setting, we need to lift the composition operations for \glspl{abk:dp} to distributions of \glspl{abk:dp}.
This is ensured by the following lemma.

\begin{lemma}\label{lem:measurable-dp-connections}
    The series~$\mthen$, parallel~$\stack$, feedback~$\trace$, union~$\union$, and intersection~$\intersection$ of \glspl{abk:dp} are measurable maps with respect to the sigma algebra generated by upper sets of \glspl{abk:dp}.
\end{lemma}

\begin{proposition}\label{cor:dp-connections-uncertain}
    The composition operations of \cref{def:dp-interconnections} can be lifted to operations between distributions of \glspl{abk:dp}.
    For two distributions of \glspl{abk:dp} $\measureP$ and $\measureQ$, whose functionality and resource \glspl{abk:poset} are compatible with the operation to be lifted, the lifted operations are
    \begin{align*}
        (\measureP \mathbin{\DistriSty{\diamond}} \measureQ) (Y) &\defeq (\measureP \measureProduct \measureQ)(\{ \tup{\dprba,\dprbb} \mid \dprba \diamond \dprbb \in Y \}), \\
        \DistriSty{\trace}(\measureP)(Y) &\defeq  \measureP(\{\dprb \mid \trace(\dprb) \in Y\}),
    \end{align*} 
    where $\diamond$ is any one of the binary operations $\mthen$, $\stack$, $\union$ or $\intersection$, and $\measureP \measureProduct \measureQ$ denotes the independent product of distributions.
\end{proposition}

Finally, we note that both interval and distributional uncertainty can be treated uniformly using the categorical structure of \emph{symmetric monoidal monads}, which captures their common structural properties through a concise set of conditions~\cite{fritzSyntheticApproachMarkov2020}.
This approach is taken in~\cite{furter2025composableuncertaintysymmetricmonoidal}, which additionally discusses uncertainties represented by subsets, widely used in robust control.


\subsection{Parameterization of \glspl{abk:dp}}\label{subsec:param-dp}

In \cref{sec:co-design}, we used a set of implementations~$\SetI$ to denote available choices when designing a component.
Each implementation $\impi$ maps to a \gls{abk:dp} that requires at least resource~$\req(\impi)$ and provides at most functionality~$\prov(\impi)$.
However, in practice, the relationship between design choices and component performance is often more nuanced.

Returning to the perception system of \glspl{abk:uav}, the key design decision is the choice of sensor and algorithm~\cite{zardiniCoDesignComplexSystems2023, milojevic2025codei} (see \cref{subsec:co-design-intro}).
Each sensor-algorithm pair yields a different \FI{detection accuracy} under varying \RI{weather conditions} and \RI{computation power}.
Hence, even for a fixed implementation, there is a dependency between functionalities and resources that cannot be captured by the \glspl{abk:mdpi} of \cref{def:mdpi}.

In addition, design choices may simultaneously influence several components in conflicting ways.
In the design of \glspl{abk:uav}, the choice of material affects the self-weight.
Lower weight benefits hardware design by reducing the required actuation force, while higher weight benefits controller performance by reducing disturbances.

These issues highlight the need for a more expressive framework that captures dependencies between component performance, design choices, and external factors that is not constrained by monotonicity requirements.
We address this by introducing \emph{parametrized \glspl{abk:dp}}.
Although our motivation stems from modeling complex performance dependencies, parametrization offers broader utility.
For instance, it enables sensitivity analysis of \glspl{abk:dp}, helping answer questions such as \emph{how much can one improve system performance by improving individual components?}
These and other directions are left for future work.

\begin{definition}[Parameterized \glspl{abk:dp}]\label{def:param-dp}
    Given a set~$\SetA$, the set of maps from $\SetA$ to $\dpOf{\funPosetF}{\resPosetR}$, denoted $\SetOfMap{\SetA}{\dpOf{\funPosetF}{\resPosetR}}$, is called \emph{\glspl{abk:dp} from $\funPosetF$ to $\resPosetR$ parametrized by $\SetA$}.
\end{definition}

We can lift operations on \glspl{abk:dp} to parameterized \glspl{abk:dp} by applying each operation element-wise.
For instance, series $
\mthen \colon \dpOf{\posetP}{\posetQ} \setproduct \dpOf{\posetQ}{\posetR} \to \dpOf{\posetP}{\posetR}$
can be lifted to 
\[
\mthenPara \colon \SetOfMap{\SetA}{\dpOf{\posetP}{\posetQ}} \setproduct \SetOfMap{\SetB}{\dpOf{\posetQ}{\posetR}} \to \SetOfMap{\SetA \setproduct \SetB}{\dpOf{\posetP}{\posetR}}, 
\]
by sending maps $\mapa \in \SetOfMap{\SetA}{\dpOf{\posetP}{\posetQ}}$ and $\mapb \in \SetOfMap{\SetB}{\dpOf{\posetQ}{\posetR}}$ to
\begin{equation*}
    \begin{aligned}
        \mapa \mthenPara \mapb \colon \SetA \setproduct \SetB  \overset{\mapa \times \mapb}{\longrightarrow} \dpOf{\posetP}{\posetQ} \times \dpOf{\posetQ}{\posetR} \overset{\mthen}{\to} \dpOf{\posetP}{\posetR}, \\
        \tup{\setelmA, \setelmB} \mapsto \tup{\mapa(\setelmA), \mapb(\setelmB)} \mapsto \mapa(\setelmA) \mthen \mapb(\setelmB).
    \end{aligned}
\end{equation*}
The remaining composition operations can be lifted similarly.
%
Furthermore, parameterized \glspl{abk:dp} can be \emph{re-parametrized} along maps with matching co-domain.

\begin{definition}[Re-parameterization of \glspl{abk:dp}]
    Given a \gls{abk:dp} $\mapa \colon \SetA \to \dpOf{\funPosetF}{\resPosetR}$ parametrized by $\SetA$ and map $\mapr \colon \SetB \to \SetA$, we can \emph{re-parameterize} $a$ using the composite $\mapa \mapFollowing \mapr \colon \SetB \to \dpOf{\funPosetF}{\resPosetR}$ that sends $\setelmB$ to $\mapa(\mapr(\setelmB))$, to obtain a \gls{abk:dp} parametrized by $\SetB$.
    Hence, $\mapr$ induces a map $\mapr^* \colon \SetOfMap{\SetA}{\dpOf{\funPosetF}{\resPosetR}} \to \SetOfMap{\SetB}{\dpOf{\funPosetF}{\resPosetR}}$.
\end{definition}

Re-parameterization can be used to express complex dependencies between design choices.
For example, suppose the parameters of the two parameterized \glspl{abk:dp} $\mapa \colon \SetA \to \dpOf{\posetP}{\posetQ}$ and $\mapb \colon \SetB \to \dpOf{\posetQ}{\posetR}$ collectively depend on design choices $\seteld \in \SetD$, represented as a map $\mapr \colon \SetD \to \SetA \setproduct \SetB$.
Then the re-parameterized \gls{abk:dp} $(\mapa \mthenPara \mapb) \mapFollowing \mapr \colon \SetD \to \dpOf{\funPosetF}{\resPosetR}$ captures how the composed problem depends on the design choice $\seteld$.
Moreover, it encodes that conditional on a fixed choice of $\seteld$, we are dealing with an independent series composition.
Such conditional independencies could be exploited during the solution of such problems.
Parametrized \glspl{abk:dp} thus provide a high-level interface for designers to specify systems in a way that implicitly provides this important information.

\begin{figure}[tb]
    \centering
    \includegraphics[width=0.7\columnwidth]{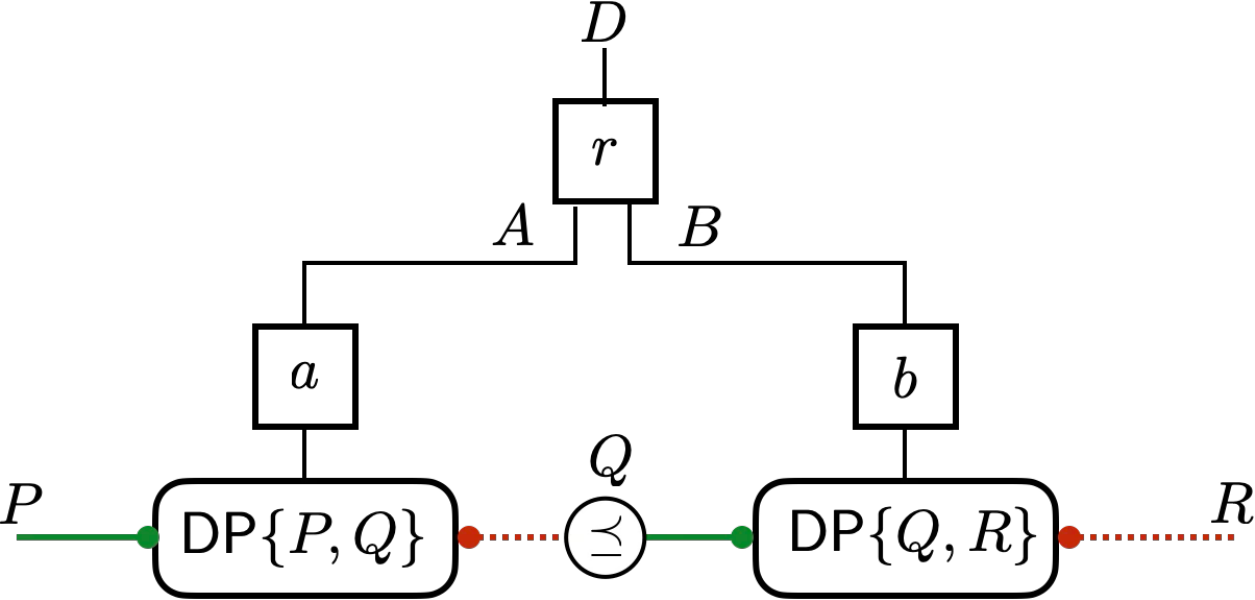}
    \caption{Series composition of parameterized \glspl{abk:dp} with re-parameterization. Rounded rectangles represent sets of design problems, and squares represent parameterization maps.}
    \label{fig:param-dp-series}
\end{figure}

We can incorporate parametrization into the diagrams used for expressing composite \glspl{abk:dp}, as shown in \cref{fig:param-dp-series}. We use additional inputs on the top of components to indicate parameter dependence and indicate re-parameterization maps using square boxes.
A more complex example of a parametric \gls{abk:dp} representing the design of \glspl{abk:uav} can be seen in  \cref{fig:uav-dp-composition-uncertain}, introduced in \cref{sec:uav-example}.

%
%


\subsection{Parameterization with uncertainty}\label{subsec:uncertain-param-dps}
\noindent We continue with the example of designing \glspl{abk:uav}, where the choice of material affects self weight, which in turn influences the performance of hardware and software components.
For a given design choice, one might further consider the uncertainty over the weight stemming from the manufacturing process, post-processing, and related factors.
This motivates the need to also introduce parametrization for uncertain \glspl{abk:dp}.
The common generalization for distributional uncertainty (\cref{subsec:uncertain-dp}) and parameterization (\cref{subsec:param-dp}) are \emph{Markov kernels} \cite{fritzSyntheticApproachMarkov2020}.

\begin{definition}[Markov kernel]\label{def:markov-kernel-distributions}
    Let $\measureSpaceA$ and $\measureSpaceB$ be \emph{measurable spaces}. A \emph{Markov kernel} $\measureKernelf \colon A \MarkovArr B$ is a map
    \[
    f \colon \SigAlgB \setproduct \SetA \to \interval{0}{1},
    \]
    satisfying the following conditions: 
    \begin{itemize}
        \item[(i)] For fixed $\setelmA \in \SetA$, the map $\measureKernelfxY{-}{\setelmA} \colon \SigAlgB \to \interval{0}{1}$ is a probability measure on $\tup{\SetB,\SigAlgB}$.
        \item[(ii)] For fixed $\subsetYB \in \SigAlgB$, the map $\measureKernelfxY{\subsetYB}{-} \colon \SetA \to \interval{0}{1}$ is measurable with respect to $\SigAlgA$.
    \end{itemize}
\end{definition}

The notation $\measureKernelfxY{\subsetYB}{\setelmA}$ emphasizes that the kernel $\measureKernelf$ can be viewed as a conditional distribution on $\SetB$, given a fixed element $\setelmA \in \SetA$.
$\measureKernelf\measureKernelApply{-}{\setelmA})$ denotes this distribution on $\SetB$.
We write $\measureKernelf: \SetA \MarkovArr \SetB$ for Markov kernels to distinguishing them from maps. Any deterministic function can be viewed as a Markov kernel.

\begin{lemma}\label{lem:markov-kernel-generalize}
    Given a measurable map $\measMapf: \SetA \to \SetB$, there is a Markov kernel $\MarkovMeasMapf: \SetA \MarkovArr \SetB$ defined by
    \begin{equation*}
        \MarkovMeasMapf\measureKernelApply{\subsetYB}{\setelmA} = \begin{cases}
            1, & \measMapf(\setelmA) \in \subsetYB, \\
            0, & \text{otherwise}.
        \end{cases}
    \end{equation*}
    sending each $\setelmA$ to the delta distribution $\deltaDistOf{\measMapf(\setelmA)}$ on $\SetB$.
\end{lemma}

As for functions, there are product and composition operations for Markov kernels.

\begin{definition}[Product of Markov kernels]\label{def:prod-markov}
    The product of measureable spaces $\measureSpaceA$ and $\measureSpaceB$ is defined as
    \begin{equation*}
        \measureSpaceA \measureProduct \measureSpaceB \defeq \tup{\SetA \setproduct \SetB, \SigAlgA \sigAlgProduct \SigAlgB},
    \end{equation*}
    where $\sigAlgProduct$ denotes the product of sigma algebras.
    Given Markov kernels $\measureKernela \colon \SetA \MarkovArr \SetB$ and $\measureKernela' \colon \SetA' \MarkovArr \SetB'$, their \emph{product} 
        $
            \measureKernela \measureProduct \measureKernela' \colon \SetA \setproduct \SetA' \MarkovArr \SetB \setproduct \SetB'
        $
    is defined by
        \[
            \measureKernelApply{\subsetYB \times Y_{\SetB'}}{\tup{\setelmA, \setelmA'}} 
            \: \mapsto \: \measureKernela\measureKernelApply{\subsetYB}{\setelmA} \: \measureKernela'\measureKernelApply{Y_{\SetB'}}{\setelmA'}. 
        \]
    Hence, for fixed $\tup{\setelmA, \setelmA'}$, the product kernel is the product of distributions.
\end{definition}

\begin{definition}[Composition of Markov kernels]\label{def:compose-markov}
    Given Markov kernels $\measureKernelf: \SetA \MarkovArr \SetB$ and $\measureKernelg: \SetB \MarkovArr \SetC$, their \emph{composite} $\measureKernelg \kernelFollowing \measureKernelf \colon \SetA \MarkovArr \SetC$ is defined as
    \begin{align*}
        \measureKernelg \kernelFollowing \measureKernelf \measureKernelApply{\subsetZC}{\setelmA} \defeq \int_{\setelmB \in \SetB} \measureKernelg \measureKernelApply{\subsetZC}{\setelmB} \measMapf \measureKernelApply{d \setelmB}{\setelmA}.
    \end{align*}
    Conceptually, $\measureKernelg \kernelFollowing \measureKernelf$ is the conditional distribution on $\SetC$, given a fixed $\setelmA \in \SetA$, obtained by marginalizing out the middle variable in $\SetB$.
\end{definition}
%



The main construction used in our case study are Markov kernels of \glspl{abk:dp} which model distributional uncertainty depending on parameters.

\begin{definition}[Uncertain parameterized \glspl{abk:dp}]\label{def:param-uncertainty-dp}
    Consider the measurable spaces $\measureSpaceA$ and $\tup{\dpOf{\funPosetF}{\resPosetR}, \sigAlgOf{\dpOf{\funPosetF}{\resPosetR}}}$. An \emph{uncertain parameterized \gls{abk:dp}} is a Markov kernel $\SetA \MarkovArr \dpOf{\funPosetF}{\resPosetR}$. It can be interpreted as a conditional distribution on $\dpOf{\funPosetF}{\resPosetR}$, conditioned on elements in $\SetA$.
\end{definition}


\begin{definition}[Re-parameterization of uncertain parameterized \glspl{abk:dp}]
    Given an uncertain parameterized \gls{abk:dp} $\measureKernela \colon \SetA \MarkovArr \dpOf{\funPosetF}{\resPosetR}$, one can re-parameterize it with a Markov kernel $\measureKernelr \colon \SetB \MarkovArr \SetA$ with matching co-domain by composing the two kernels: $\measureKernela \kernelFollowing \measureKernelr \colon \SetB \MarkovArr \dpOf{\funPosetF}{\resPosetR}$.
\end{definition}

As in the deterministic case, operations for uncertain \glspl{abk:dp} can be lifted to uncertain parameterized \glspl{abk:dp} by applying them element-wise.
For instance, given two uncertain parameterized \glspl{abk:dp}~$\measureKernela \colon \SetA \MarkovArr \dpOf{\posetP}{\posetQ}$ and $\measureKernelb \colon \SetB \MarkovArr \dpOf{\posetQ}{\posetR}$, their lifted composition $\mapa \mthenUPara \mapb$ is defined by
\begin{equation*}
    \SetA \setproduct \SetB \MarkovArrOf{\measureKernela \measureProduct \measureKernelb} 
    \dpOf{\posetP}{\posetQ} \setproduct \dpOf{\posetQ}{\posetR} \MarkovArrOf{\mthenMarkov}
    \dpOf{\posetP}{\posetR},
\end{equation*}
where~$\mthenMarkov$ is the Markov kernel lifting series composition~$\mthen$ according to \cref{lem:markov-kernel-generalize}.

Again, one may introduce dependencies between parameters by reparametrizing with a Markov kernel~$\measureKernelf \colon \SetD \MarkovArr \SetA \setproduct \SetB$, which represents a conditional distribution on~$\SetA \setproduct \SetB$, given a specific design choice in~$\SetD$.
Moreover, diagrams such as \cref{fig:param-dp-series} and \cref{fig:uav-dp-composition-uncertain} can also represent uncertain parameterized \glspl{abk:dp}, by interpreting the squares as Markov kernels.
Finally, it is possible to introduce parametric versions of interval uncertainty using analogous definitions. In fact, both of these cases can be treated uniformly using category theory~\cite{furter2025composableuncertaintysymmetricmonoidal} allowing (parametrized) co-design diagrams to be endowed with any uncertainty semantics forming a symmetric monoidal monad.

\subsection{Queries with uncertainty and parameterization}\label{subsec:queries-with-uncertainty-param}

Adding uncertainty and parametrization to co-design makes querying more nuanced. For example, consider the interpretation of $\union$, which represents freely choosing between designs.
On the one hand, we could be forced to fix our choice \emph{before} the true value of design parameters are known.
On the other, we could be allowed to make our choice \emph{after} learning the values of uncertain design parameters.

For both interval and distributional uncertainty, the lifted union operation represents the latter case. To see this, recall that by \cref{lem:interval-lifted-ops} the lifted union is
$$
\interval{\dprba}{\dprbb} \unionInterval \interval{\dprba'}{\dprbb'} = \interval{\dprba \union \dprba'}{\dprbb \union \dprbb'}.
$$
The worst- and best-case bounds on the right imply that one may select the preferable design \emph{after} encountering concrete instances from each interval on the left:
In the worst-case scenario, the left yields $\dprba$ and $\dprba'$, and we then freely choose between them. Similarly, in the best-case scenario, we freely choose between $\dprbb$ and $\dprbb'$.


For distributional uncertainty, the lifted union is given by
$$
(\measureP \mathbin{{\DistriSty{\union}}} \measureQ) (Y) = (\measureP \measureProduct \measureQ)(\{ \tup{\dprba,\dprbb} \mid \dprba \union \dprbb \in Y \}),
$$
according to \cref{cor:dp-connections-uncertain}.
This equation states that the probability of satisfying event $Y$ under the lifted choice can be calculated by independently sampling pairs $\dprba$ and $\dprbb$ from $\measureP$ and $\measureQ$, and checking if the free choice $\dprba \union \dprbb$ satisfies $Y$. Hence the feasibility of the choice is evaluated \emph{after} the specifics of the \glspl{abk:dp} are known.

In contrast, making design choices \emph{before} the true system is known is a new type of query for uncertain co-design. 
We can address it by modeling design choices as parameters for individual components, as illustrated in \cref{fig:uav-dp-composition-uncertain}. Compositionality ensures that we can interpret the resulting \gls{abk:dp} as a Markov kernel~$\measureKernelSys \colon \SetD \MarkovArr \dpOf{\funPosetF}{\resPosetR}$, with $\SetD$ being the set of design choices, $\funPosetF$ being the poset of \FI{functionalities} provided by the system, and $\resPosetR$ being the \RI{resources} required. Hence, we must pick a design choice $\seteld \in \SetD$ that yields an optimal distribution $\measureKernelSys\measureKernelApply{-}{\seteld}$ of \glspl{abk:dp}. What is considered optimal will depend on the specific application and can incorporate notions of risk.

A similar issue arises when fixing a
functionality requirement $\poselxF \in \funPosetF$ and resource budget $\poselxR \in \resPosetR$. If these are specified \emph{before} we have full knowledge of the system, we can calculate the success probability of a design choice $\seteld \in \SetD$ by $\measureKernelSys\measureKernelApply{\setWithArg{\dprba}{\dprba(\poselxF, \poselxR)}}{\seteld}$.
If we want to leave the selection of $\poselxF$ and $\poselxR$ until \emph{after} the \gls{abk:dp} is (partially) known, the design process becomes a stochastic adaptive optimization problem \cite{marti_stochastic_2024, bertsimasRobustAdaptiveOptimization2022}.

Computing the kernel~$\measureKernelSys$ involves composing Markov kernels and calculating the lifted operations for \glspl{abk:dp}, both of which rarely have closed-form solutions.
Moreover, the domain of $\measureKernelSys$ is a large, non-numerical space of design problems~$\dpOf{\funPosetF}{\resPosetR}$.
Developing general and efficient methods for composing and querying uncertain co-design problems remains an open direction for future work.



\section{Uncertain Task-driven \gls{abk:uav} Co-design}\label{sec:uav-example}
\noindent In this section, we apply our theory to the uncertainty-aware task-driven co-design of \glspl{abk:uav}.
The problem is to design a \gls{abk:uav} that completes delivery tasks (adapted from~\cite{censi2017uncertainty}).
We first show how one can decompose the \gls{abk:dp} into components and how the problem can be solved in the deterministic case.
Then, we introduce distributional uncertainties into some components and show how they fit into the framework introduced in \cref{sec:uncertain-param-dp}.
Finally, we estimate the uncertain \gls{abk:dp} via Monte-Carlo sampling, illustrating how the distributional uncertainties in \glspl{abk:dp} propagate to the final trade-offs that interest system designers. 

\subsection{Task-driven co-design of \glspl{abk:uav}}\label{subsec:deterministic-uav}
\begin{figure}[tb]
    \centering
    \includegraphics[width=\columnwidth]{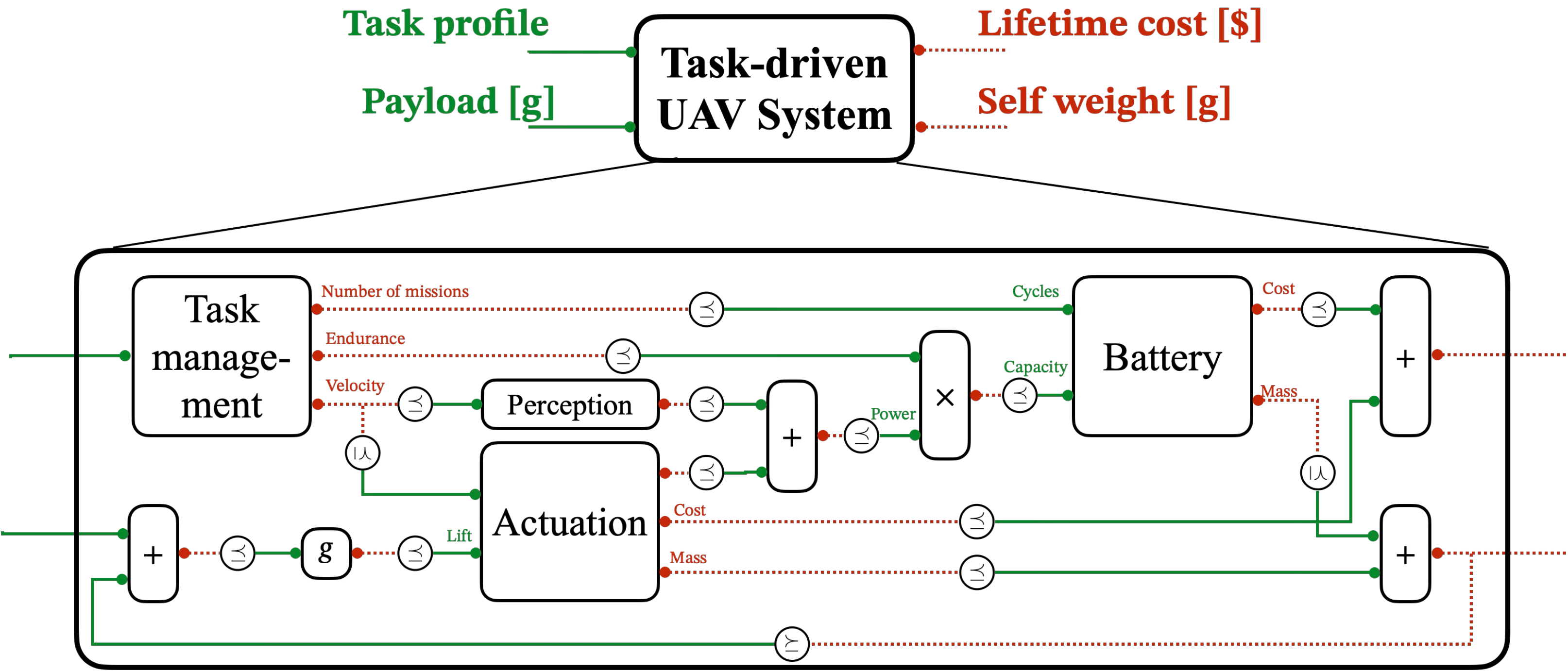}
    \caption{Co-design diagram for a task-driven \gls{abk:uav}, showing how the system can be decomposed into functional components.}
    \label{fig:uav-dp-interface}
    \label{fig:uav-dp-composition}
\end{figure}

\noindent Suppose we aim to design a \gls{abk:uav} capable of completing package delivery tasks within its operational lifetime.
The \FI{task profile} is specified by three key requirements: \FI{number of delivery missions}, \FI{distance coverage}, and \FI{mission frequency}.
Given a specific task profile, our objective is to determine the optimal design that maximizes the \FI{payload} capacity of the \gls{abk:uav} while minimizing both its total \RI{lifetime cost} and \RI{self weight}.
Within the co-design framework, this objective is captured by a \gls{abk:dp} that provides the functionalities \FI{task profile} and \FI{payload}, and requires the resources \RI{lifetime cost} and \RI{self weight} (\cref{fig:uav-dp-composition}, top).
Importantly, our framework allows one to further decompose the design into sub-systems (\cref{fig:uav-dp-composition}, bottom) by leveraging functional decomposition~\cite{zardiniCoDesignComplexSystems2023}.
The decomposed diagram highlights essential components involved in task-driven \gls{abk:uav} co-design, namely task management, perception, actuation, and battery.
For clarity, the system architecture and sub-system models are simplified, with design choices limited to \emph{battery types} and \emph{actuators}.
Our co-design framework readily allows for more detailed models with minimal overhead (e.g.,~\cite{milojevic2025codei}).
\paragraph*{Task management}
Derives the specifications of the \gls{abk:uav}, including \RI{number of missions} required, \RI{endurance} (battery life), and \RI{velocity}, based on a given \FI{task profile}. 
\paragraph*{Perception}
We assume the perception system (sensor and software), is provided and fixed.
However, it consumes more \RI{power} as the \FI{velocity} of the \gls{abk:uav} increases~\cite{karaman2012high}.
\paragraph*{Actuation}
For the actuation system, we assume a choice among three different motors, each characterized by a specific \RI{cost} and \RI{mass}, and offering a defined maximum \FI{velocity}.
The \RI{power} consumption~$P$ of each motor is modeled as a function of the required \FI{lift} $F$: $P \geq p_0 + p_1 \times F^2$, where $p_0$ and $p_1$ are motor-specific parameters (\cref{tab:param-actuators}).
\paragraph*{Battery}
The most critical design choice for the battery system is the selection of technology, defined by three parameters: \emph{power density}, \emph{cost per unit power}, and the \emph{number of operating cycles before maintenance}.
Given a technology, the system provides \FI{capacity} and a number of \FI{cycles} considering replacements at the expense of \RI{mass} and \RI{total cost}.
The total cost accounts for both initial purchase and maintenance/replacement expenses.
The parameters for different battery technologies are listed in \cref{tab:param-battery-types}.

\begin{table}[tb]
    \centering
    \begin{tabular}{l C{0.7cm} C{0.7cm} C{0.7cm} C{0.7cm} C{0.7cm}}
        \hline\hline
        {Actuator} & {Mass} [\si{g}] & {Cost} \newline [\si{\$}] & {Velocity} \newline [\si{m/s}] & $p_0$ \newline [\si{W}] & $p_1$ \newline [\si{W/N^2}] \\
        \hline 
         $a_1$ & 50.0 & 50.0 & 3.0 & 1.0 & 2.0 \\
         $a_2$ & 100.0 & 100.0 & 3.0 & 2.0 & 1.5 \\
         $a_3$ & 150.0 & 150.0 & 3.0 & 3.0 & 1.5 \\
         \hline\hline
    \end{tabular}
    \caption{Deterministic parameters of actuators.}
    \label{tab:param-actuators}
\end{table}

\begin{table}[tb]
    \centering
    \begin{tabular}{l C{1.6cm} C{1.6cm} C{1.6cm}}
        \hline\hline
        {Technology} & {Energy density} [\si{Wh/kg}] & {Unit power per cost} [\si{Wh/\$}] & {Number of cycles} \\
        \hline 
         NiMH & 100.0 & 3.41 & 500 \\
         NiH2 & 45.0 & 10.50 & 20,000 \\
         LCO & 195.0 & 2.84 & 750 \\
         LMO & 150.0 & 2.84 & 500 \\
         NiCad & 30.0 & 7.50 & 500 \\
         SLA & 30.0 & 7.00 & 500 \\
         LiPo & 150.0 & 2.50 & 600 \\
         LFP & 90.0 & 1.50 & 1,500 \\
         \hline\hline
    \end{tabular}
    \caption{Deterministic parameters of battery technologies~\cite{censi2017uncertainty}.}
    \label{tab:param-battery-types}
\end{table}

\begin{figure}[tb]
    \centering
    \includegraphics[width=\columnwidth]{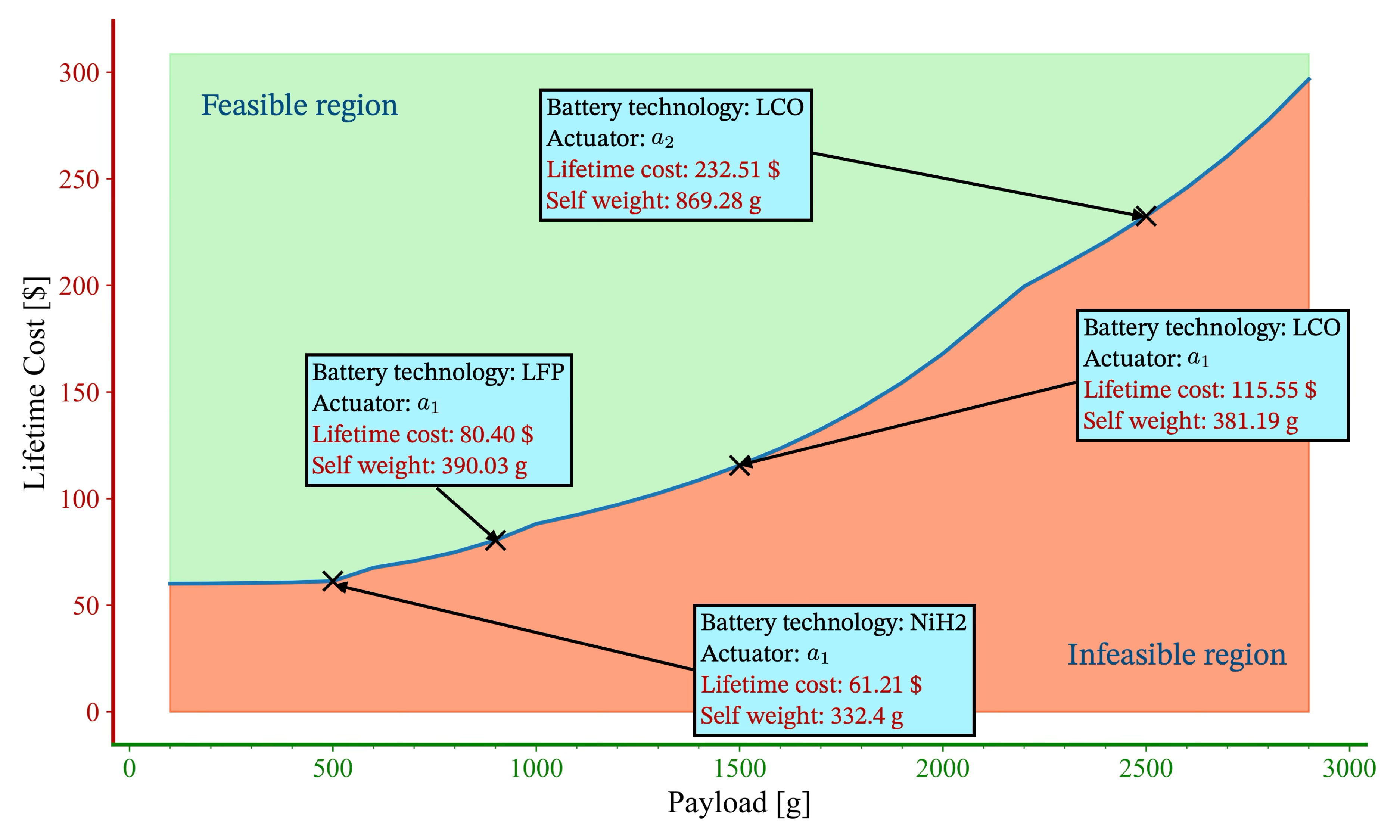}
    \caption{Trade-off between \FI{payload} and \RI{lifetime cost} for a fixed \FI{task profile}, free choice of battery and actuators, and deterministic battery and actuator parameters. Pairs of \FI{payload} and \RI{lifetime cost} in the green region are feasible while those in the orange region are not. A larger \FI{payload} requires higher \RI{lifetime cost}, illustrating the monotonicity of co-design. Specific choices of battery technologies and actuators are shown for some optimal solutions.}
    \label{fig:deterministic-trade-off}
\end{figure}

Solving the deterministic co-design problem with fixed \FI{task profile}, a free choice of battery technologies and actuators, and only optimizing for \RI{lifetime cost}, yields the trade-off between \FI{payload} and \RI{lifetime cost} shown in \cref{fig:deterministic-trade-off}.

\subsection{Uncertainty in battery and actuation parameters}\label{subsec:uncertain-uav}
\begin{figure}[tb]
    \centering
    \includegraphics[width=\columnwidth]{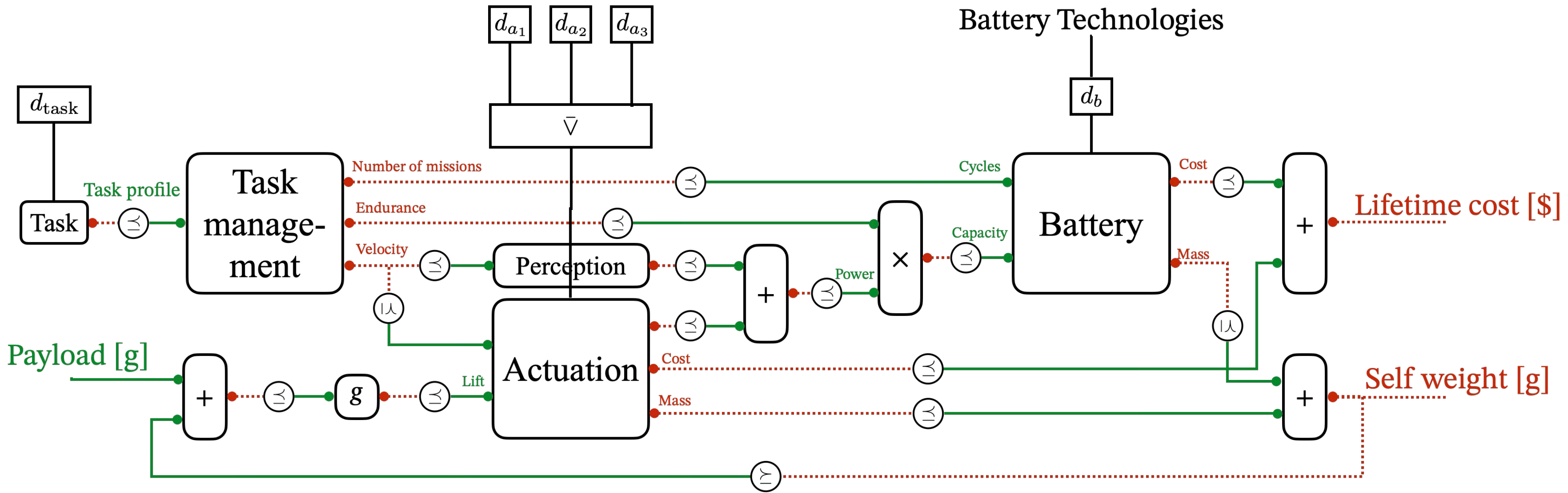}
    \caption{Co-design diagram for the uncertainty-aware task-driven co-design of a \gls{abk:uav}. The square boxes represent parameterized uncertainties implemented by Markov kernels.}
    \label{fig:uav-dp-composition-uncertain}
\end{figure}

We now introduce uncertainty into the \gls{abk:uav} co-design problem and present numerical results.
Specifically, we consider uncertainty in the task profile and the battery and actuation parameters. 
The resulting co-design problem with parametrized uncertainty is depicted in \cref{fig:uav-dp-composition-uncertain}. 
We include a \gls{abk:dp}~$\dprbtask$ that provides no functionalities but requires satisfaction of a given \FI{task profile}.
This setup enables us to consider probability distributions over the \FI{task profile} by leveraging the framework of \cref{sec:uncertain-param-dp}. Such distributions might represent partial information about the task, or how the task is expected to vary over the \gls{abk:uav}'s lifetime.

We assume that the actuator parameters \emph{cost} and \emph{mass} remain deterministic.
However, other parameters, such as energy density and the coefficient~$p_1$, are modeled as Gaussian random variables with mean equal to the deterministic value and variance calibrated so that the boundary of \SI{90}{\percent} confidence interval corresponds to \SI{\pm 5}{\percent} deviation from the mean value. This could represent our ignorance of the deterministic parameter value, or some random variation that occurs during manufacturing.

We assume that we can choose the actuator \emph{after} observing the true parameters of the actuator component.
As discussed in \cref{subsec:queries-with-uncertainty-param}, this scenario corresponds to the lifted union operation, which is included as a re-parametrization box in \cref{fig:uav-dp-composition-uncertain}.
On the other hand, we assume the battery technology must be selected \emph{before} observing battery parameters. This might reflect real-world constraints where manufacturing decisions precede the full characterization of component behavior. As explained in \cref{subsec:queries-with-uncertainty-param}, we model such post-sampling choices using external parameters.
The composite system of \cref{fig:uav-dp-composition-uncertain} represents a Markov kernel~$\measureKernelUAV \colon \SetTB \MarkovArr \dpOf{\F{\posReals}}{\R{\posReals^2}}$, where
$\SetTB$ is a finite set of battery technologies, the functionality poset represents \FI{payload}, and the resource poset represents \RI{self weight} and \RI{lifetime cost}.

Suppose we want to minimize \RI{lifetime cost}, ignoring \RI{self weight}.
For each $\seteltB \in \SetTB$, we use Monte-Carlo sampling to obtain samples from the distribution~$\measureKernelUAV\measureKernelApply{-}{\seteltB}$.
The lifted union operator~$\unionDistri$ can be computed by applying union~$\union$ to samples. We then push each samples from~$\measureKernelUAV\measureKernelApply{-}{\seteltB}$ through the deterministic co-design solver \cite{zardiniCoDesignComplexSystems2023}: For each target \FI{payload}, we obtain the minimal \RI{lifetime cost} for the sampled \gls{abk:dp}.
The resulting samples approximate the distribution of minimal \RI{lifetime cost} that we expect technology $\seteltB$ to require for reaching the target \FI{payload}.

\begin{figure}[tb]
    \centering
    \includegraphics[width=\columnwidth]{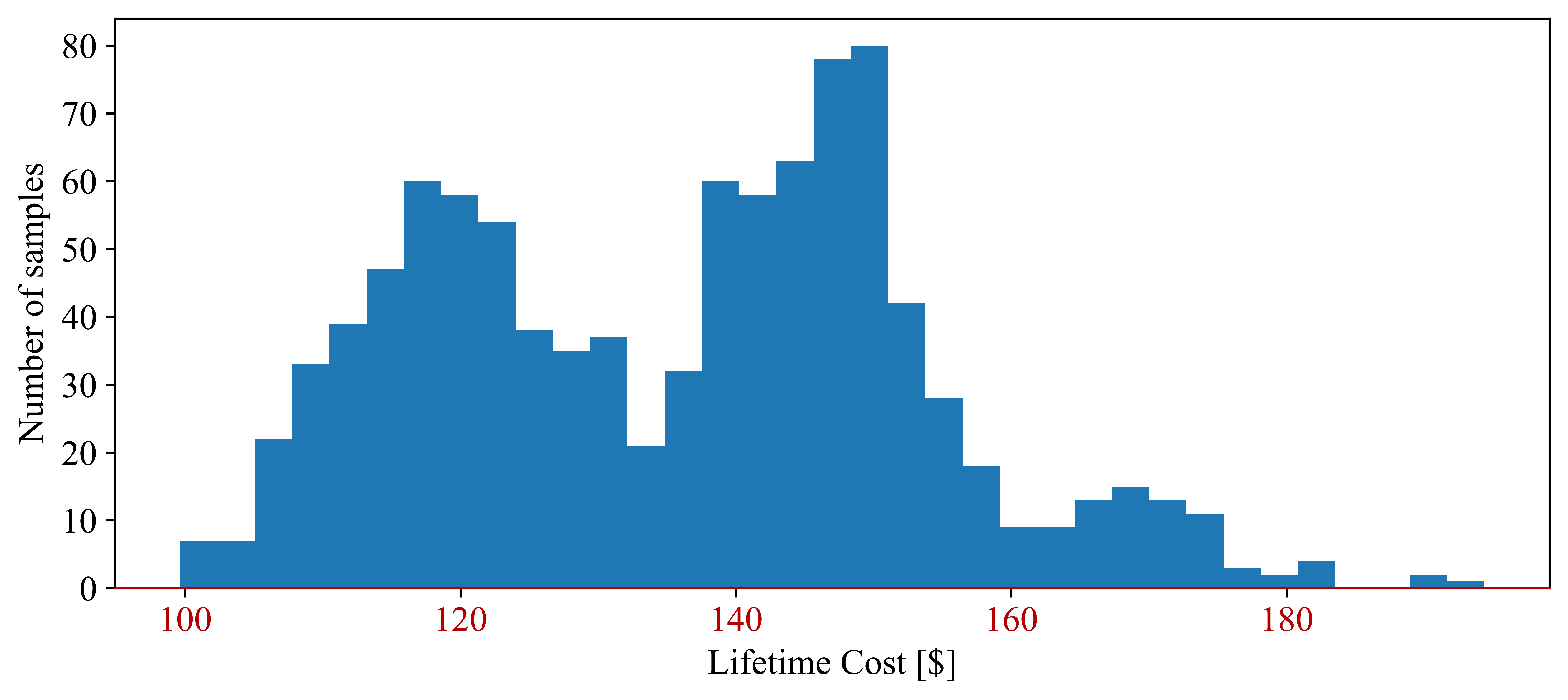}
    \caption{Histogram of minimal \RI{lifetime cost} requirements for the uncertainty-aware co-design of a \gls{abk:uav} with $\text{NiMH}$ battery and \SI{1300}{g} \FI{payload}.}
    \label{fig:samples-one-battery-payload}
\end{figure}

\begin{figure}[tb]
\begin{center}
\begin{subfigure}[b]{0.49\columnwidth}
    \centering
    \includegraphics[width=\textwidth]{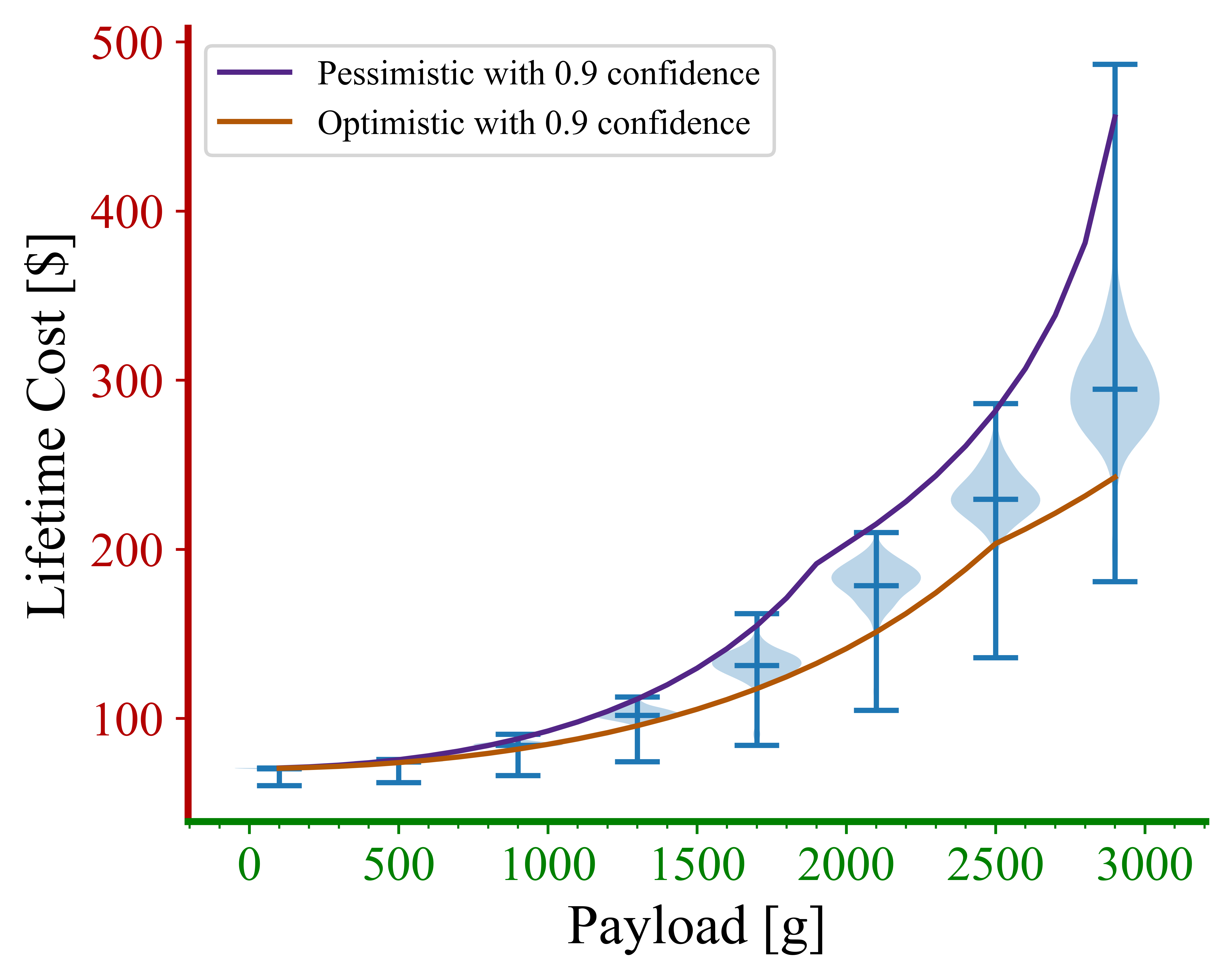}
    \subcaption{Battery technology: LCO.}
\end{subfigure}
\begin{subfigure}[b]{0.49\columnwidth}
    \centering
    \includegraphics[width=\textwidth]{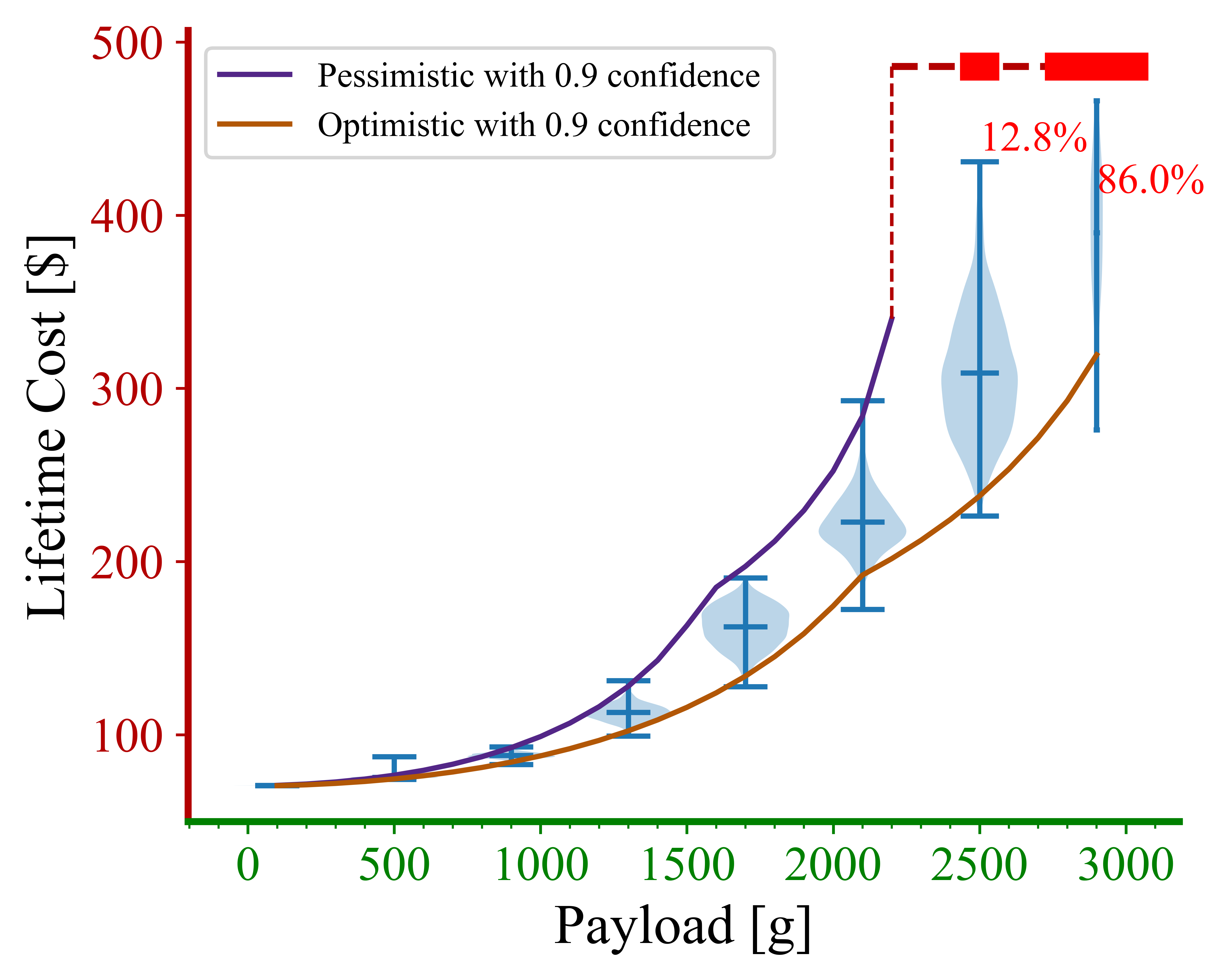}
    \subcaption{Battery technology: LiPo.}
\end{subfigure}
\begin{subfigure}[b]{0.49\columnwidth}
    \centering
    \includegraphics[width=\textwidth]{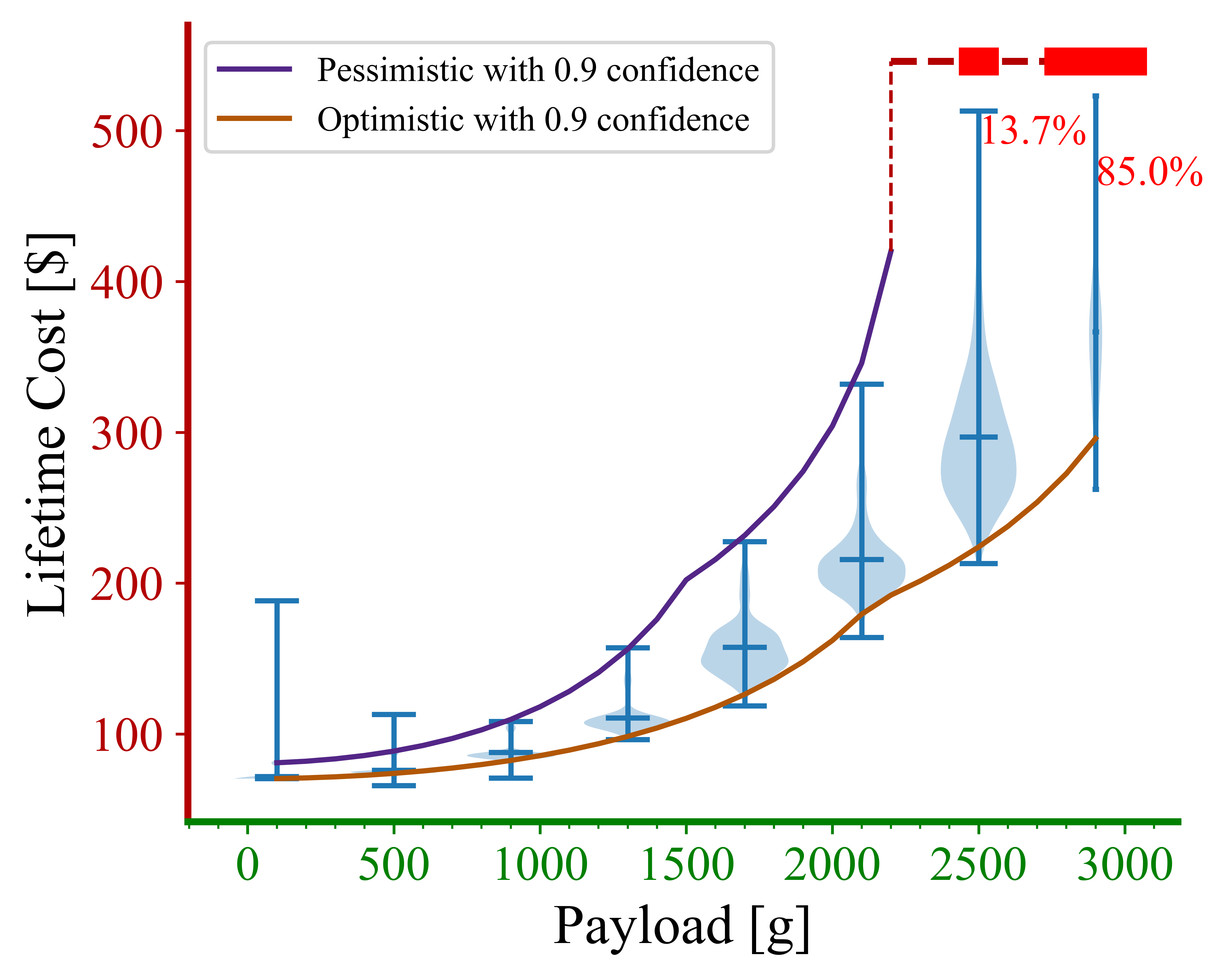}
    \subcaption{Battery technology: LMO.}
\end{subfigure}
\begin{subfigure}[b]{0.49\columnwidth}
    \centering
    \includegraphics[width=\textwidth]{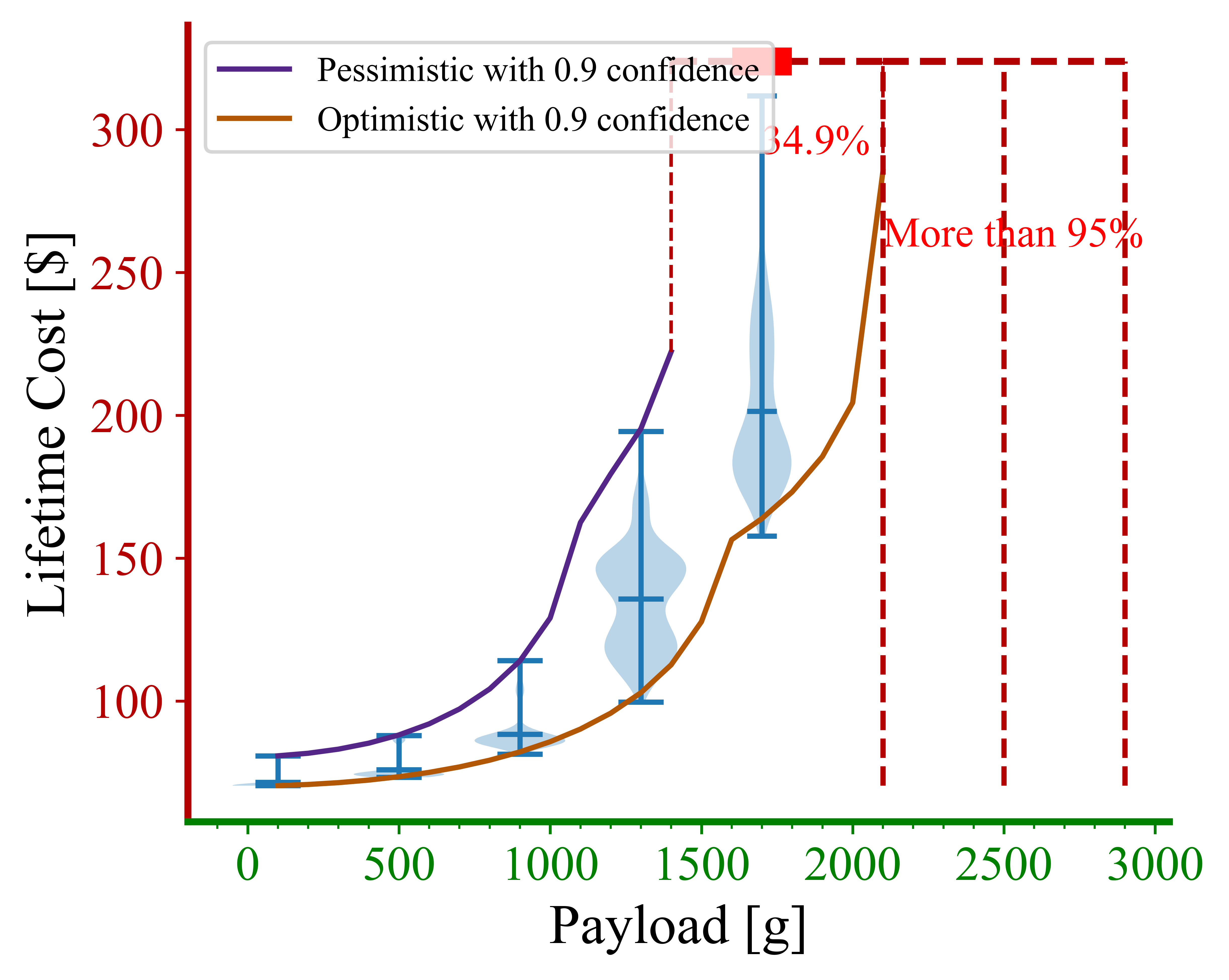}
    \subcaption{Battery technology: NiMH.}
\end{subfigure}
\caption{Trade-offs for \RI{lifetime cost} ($y$-axis) resulting from the uncertainty-aware co-design of a \gls{abk:uav} with varying target \FI{payloads} ($x$-axis) and battery technologies (panels (a)-(d)). For fixed target \FI{payload}, the violin plot above it shows the distribution of minimal \RI{lifetime cost} that one can expect from the design. Solid lines represent the optimistic and pessimistic bounds: For fixed \FI{payload}, the probability that \RI{lifetime cost} lies below the blue line is \SI{95}{\percent}, while the probability of it exceeding the orange line is \SI{95}{\percent}. Red text indicates the percentage of samples that are infeasible under any cost. Each distribution was estimated from 1,000 samples.}
\label{fig:violin-plot-batteries}
\end{center}
\vspace{-0.5cm}
\end{figure}

%
\Cref{fig:samples-one-battery-payload} shows a histogram of such minimal costs for $\text{NiMH}$ battery and \SI{1300}{g} \FI{payload}.
Although the parameter distributions are all Gaussian, this distribution has multiple modes. Plots illustrating how this distribution varies with different battery technologies and payloads are displayed in \cref{fig:violin-plot-batteries}. We observe that different technologies have distinct advantages.
For instance, at low \FI{payload}s, NiMH provides the best expected \RI{lifetime cost}, albeit with a larger variance than other technologies. Such insights would be impossible without accounting for uncertainty. Moreover, distributional information allows for more precision than worst- and best-case bounds. For LiPo at a \FI{payload} of \SI{1750}{g}, the design is infeasible at worst and costs under \SI{200}{\$} at best. However, our more detailed analysis allows us to narrow down this large range to its probable values. For example, it shows that designs are infeasible with a probability of \SI{12.8}{\percent}. Nonetheless, carrying forward precise distributional information introduces challenging stochastic multi-objective optimization problems.


\section{Conclusions}
\label{sec:conclusions}
We presented a unified compositional framework for incorporating uncertainty into monotone co-design, extending the classic formulation to handle interval, distributional, and parameterized uncertainty.
By lifting co-design operations to these richer structures, we enabled expressive and modular reasoning about trade-offs under uncertainty.
This was illustrated in a case study on uncertainty-aware \gls{abk:uav} co-design that showcased the practical relevance of the framework in capturing both design flexibility and risk.
Promising future directions include estimating parameterized uncertain \glspl{abk:dp} from data, developing efficient algorithms for specific optimization queries, and applying stochastic, adaptive optimization methods to this framework.

\appendix
\label{sec:appendix}

\subsection{Proof sketches} \label{sec:proofs}
\label{sec:proof_sketches}

\begin{proof}[\cref{lem:inclusion-to-dist}]
    One directly checks that $\delta(\poselxP)$ and $\DistriOf{f}(\measureP)$ satisfy the axioms of a probability distribution.
\end{proof}

\begin{proof}[\cref{lem:measurable-dp-connections}]
    We begin by noting that all the composition operations in the lemma are monotone maps.
    Moreover, any monotone map $\mapf \colon \posetP \to \posetQ$ lifts to a measurable map $\mapf' \colon \tup{\posetP, \sigAlgOf{\posetP}} \to \tup{\posetQ, \sigAlgOf{\posetQ}}$ with the same underlying function, since preimages of upper sets under $f$ are again upper sets, showing that $\mapf$ is measurable on generators. This is sufficient to show that the unary trace operator $\trace$ is measurable. The binary operations are monotone maps of the form $\diamond \colon \posetP_1 \times \posetP_2 \to \posetQ$ and hence lift to measurable maps $\diamond' \colon \tup{\posetP_1 \times \posetP_2, \sigAlgOf{\posetP_1 \times \posetP_2}} \to \tup{\posetQ, \sigAlgOf{\posetQ}}$.
\end{proof}

\begin{proof}[\cref{cor:dp-connections-uncertain}]
By \cref{lem:measurable-dp-connections}, trace is a measurable map $\trace \colon \tup{\posetP, \sigAlgOf{\posetP}} \to \tup{\posetQ, \sigAlgOf{\posetQ}}$ and hence lifts to a map $\hat \trace := \DistriOf{\trace} \colon \DistriOf{\posetP} \to \DistriOf{\posetQ}$. Similarly, binary operations $\mthen$, $\stack$,  $\union$ and $\intersection$ are measurable maps of the form $\diamond \colon \tup{\posetP_1 \times \posetP_2, \sigAlgOf{\posetP_1 \times \posetP_2}} \to \tup{\posetQ, \sigAlgOf{\posetQ}}$ and thus lift to maps $\DistriOf{\diamond} \colon \DistriOf{\posetP_1 \times \posetP_2} \to \DistriOf{\posetQ}$. 
    Let $X$ denote the set of probability distributions on the product space $\tup{\posetP_1 \times \posetP_2, \sigAlgOf{\posetP} \otimes \sigAlgOf{\posetQ} }$, where $\sigAlgOf{\posetP} \otimes \sigAlgOf{\posetQ}$ is the product sigma algebra. Using basic measure theory and facts about upper sets one can show that $X = \DistriOf{\posetP_1 \times \posetP_2}$ \cite[Appendix E]{furter2025composableuncertaintysymmetricmonoidal}. Hence, we can precompose $\DistriOf{\diamond}$ with the map $\pi: \DistriOf{\posetP_1} \times \DistriOf{\posetP_2} \to X = \DistriOf{\posetP_1 \times \posetP_2}$ that sends a pair of distributions to their product distribution to obtain the desired lifted operations.
\end{proof}

\bibliographystyle{IEEEtran}
\bibliography{paper}

\end{document}